\documentclass[usenatbib]{mnras}

\usepackage{amsmath}
\usepackage{amssymb}	% Extra maths symbols
\usepackage{graphicx}
\usepackage[T1]{fontenc}
\usepackage{ae,aecompl}

\title[Quenching star formation using AGN feedback]{Cosmic evolution of stellar quenching by AGN feedback: clues from the Horizon-AGN simulation}
\author[R. S. Beckmann et al.]{R. S. Beckmann$^{1,}$\thanks{Email: ricarda.beckmann@physics.ox.ac.uk}, J. Devriendt$^{1,2}$, A. Slyz$^{1}$, S. Peirani$^{3,4}$, M. L. A. Richardson$^{1}$,\newauthor Y. Dubois$^{3,4}$,  C. Pichon$^{3,4,5}$, N. E. Chisari$^{1}$, S. Kaviraj$^{6}$, C. Laigle$^{1}$, M. Volonteri$^{3,4}$
\\ \\
$^{1}$ Sub-department of Astrophysics, University of Oxford, Keble Road, Oxford OX1 3RH, UK\\
$^{2}$ Observatoire de Lyon, UMR 5574, 9 avenue Charles Andre, F-69561, Saint Genis Laval , France \\
$^{3}$ Sorbonne Universites, UPMC Univ Paris 06, UMR 7095, Institut d'Astrophysique de Paris, F-75005, Paris, France\\
$^{4}$ CNRS, UMR 7095, Institut d'Astrophysique de Paris, 98bis Boulevard Arago, F-75014, Paris, France \\
$^{5}$ Korea Institute for Advanced Study (KIAS), 85 Hoegiro, Dongdaemun-gu, Seoul, 02455, Republic of Korea \\
$^{6}$ Centre for Astrophysics Research, University of Hertfordshire, College Lane, Hatfield, Herts, AL10 9AB}
\pubyear{2017}

\begin{document}
\label{firstpage}	
\pagerange{\pageref{firstpage}--\pageref{lastpage}}
\maketitle

\begin{abstract}
The observed massive end of the galaxy stellar mass function is steeper than its predicted dark matter halo counterpart in the standard \(\Lambda \)CDM paradigm. In this paper, we investigate the impact of active galactic nuclei (AGN) feedback on star formation in massive galaxies. We isolate the impact of AGNs by comparing two simulations from the HORIZON suite, which are identical except that one also includes super massive black holes (SMBH), and related feedback models. This allows us to cross-identify individual galaxies between simulations and quantify the effect of AGN feedback on their properties, including stellar mass and gas outflows. We find that massive galaxies (\( \rm M_{*} \geq 10^{11} M_\odot  \)) are quenched by AGN feedback to the extent that their stellar masses decrease by up to 80\% at $z=0$. SMBHs affect their host halo through a combination of outflows that reduce their baryonic mass, particularly for galaxies in the mass range \( \rm 10^9 M_\odot \leq M_{*} \leq  10^{11} M_\odot  \), and a disruption of central gas inflows, which limits in-situ star formation. As a result, net gas inflows onto massive galaxies,  \( \rm M_{*} \geq 10^{11} M_\odot  \), drop by up to 70\%. We measure a redshift evolution in the stellar mass ratio of twin galaxies with and without AGN feedback, with galaxies of a given stellar mass showing stronger signs of quenching earlier on. This evolution is driven by a progressive flattening of the \(\rm M_{\rm SMBH}-M_* \) relation with redshift, particularly for galaxies with \(\rm M_{*} \leq  10^{10} M_\odot  \). $\rm M_{\rm SMBH}/M_*$ ratios decrease over time, as falling average gas densities in galaxies curb SMBH growth.
\end{abstract}

\begin{keywords}
galaxies: evolution - galaxies: high-redshift - galaxies: quasars: supermassive black holes - galaxies: star formation - galaxies: active -  methods: numerical \\
\end{keywords}

%%%%%%%%%%%%%%%%%%%%%%%%%%%%

\section{Introduction}
It has long been known that the hierarchical structure formation paradigm implied by the cold dark matter model, while very successful overall, overproduces objects at the bright and faint end of the luminosity function \citep{White1991,Kauffmann1993}.  Observations show much more inefficient star formation in low and high mass halos, with a peak in efficiency at the luminosity turnover \citep{Bernardi2013,Moustakas2013,Davidzon2017}. To avoid the overcooling problem, and reproduce the observed luminosity function, an energetic feedback mechanism is required \citep{Cole1994, Binney1995, Silk1998, Springel2005, Dubois2013}. Several different avenues have been suggested to provide the necessary energy input and quench star formation, including the extragalactic UV background, supernova (SN) feedback and feedback due to active galactic nuclei (AGN). 

Photoionisation by the extragalactic UV background and the first generations of stars suppresses gas accretion at high redshift, causing a number of smaller dark matter (DM) halos to remain devoid of gas \citep{Gnedin2000,Somerville2002,Okamoto2008,Geen2013}. While this mechanism provides a possible solution to the overabundance of very low mass ($\leq 10^8$M$_\odot$) substructures in Milky-Way like halos, it has been shown to have little effect on more massive objects that collapse later \citep{Efstathiou2000,Governato2007}.

On the other hand, SN feedback is widely believed to play an important part in quenching star formation in halos with masses below $10^{11}$ M$_\odot$ \citep{Dekel1986,Benson2003,Hopkins2014}. Their shallow potentials allow SN driven winds with velocities comparable to the escape velocity \citep{MacLow1999,Christensen2016,Keller2016} to empty the host galaxy of a significant amount of gas, thereby efficiently suppressing star formation. However, even stellar feedback concentrated in intense, compact starbursts caused by major mergers or violent disc instabilities cannot quench massive galaxies \citep{Dekel2014}. 

It has been suggested that star formation halts in massive objects due to a slowdown in cold flows at low redshift \citep{Feldmann2014}. However \citet{Mandelbaum2015} show that massive quiescent galaxies can have twice as much DM as star forming galaxies, indicating that cosmic inflows probably continue long after star formation has ceased. Furthermore, the slowdown in cold flows is expected to take place over Gyr timescales \citep{Zolotov2015}, which contradicts observational evidence that quenching of massive galaxies takes place on much shorter timescales \citep{Barro2015,Tomczak2016}. Therefore the reduction of star formation in massive galaxies is unlikely to occur solely because accretion is fading away.

Instead, AGN feedback could provide an effective route to quenching massive galaxies, as well as regulating the growth of supermassive black holes (SMBH) \citep{Binney1995,Benson2003,DiMatteo2005,Bower2006,Croton2006,Sijacki2007,Cattaneo2009,Fabian2012}. There are two main mechanisms through which this could proceed. One possibility is that black holes and stars feed from the same cold gas supply until it is depleted by AGN feedback, at which point both processes come to a halt. Observations of galaxies that simultaneously harbour both an AGN and an active starburst provide evidence that supports this claim \citep{Heckman2006,Nardini2008,Lutz2008}. In this scenario, AGN feedback might even accelerate star formation by further compressing the cold gas of the galaxy, in a so called positive feedback mode \citep{Silk2005,Gaibler2012,Santini2012, Bieri2015}. However, its main
role is to prevent the gas heated/expelled by SN winds to be re-accreted at a later stage, alongside more pristine material. This is the so called maintenance mode, associated with powerful radio emission \citep{Rafferty2006,Cattaneo2007}.

Alternatively, AGN feedback could act directly on the gas content of the galaxy. It could expel the interstellar medium (ISM) out of galaxies in massive galactic winds, and/or prevent star formation by directly heating the ISM gas \citep{Springel2005,DiMatteo2005,Murray2005,Fabian2006}. This view is supported by observational evidence of frequent and fast outflows in massive galaxies \citep{Tremonti2007}, able to drive a significant gas mass \citep{Heckman2000,Veilleux2005,Weiner2009,Sturm2011} using only \(5-10\% \) of accretion power \citep{Moe2009,Saez2009,Dunn2010}. Whilst such outflows are common in AGN with powerful radio jets \citep{DeKool2000,Arav2001,Reeves2003,Chartas2007,Dunn2010,Moe2009}, careful analysis of higher redshift objects provides evidence that quasars can also launch powerful energy driven winds and thus cause a rapid star formation decline \citep{Maiolino2012,Page2012,Farrah2012,Cicone2014,Costa2014a,Williams2016}. Both modes of AGN feedback can reproduce the observed correlations between host galaxy and BH properties \citep{Ferrarese2000,Gebhardt2000}, as shown by e.g. \cite{DiMatteo2008,Booth2009,Dubois2012,Sijacki2015,Volonteri2016}. However, the timescale over which quenching takes place is still a matter of debate, with evidence existing for both rapid quenching \citep{Thomas2005} and much slower processes \citep{Quintero2004}. The timescales are probably dependent on galaxy type \citep{Schawinski2014}.

In this work we use state-of-the-art cosmological simulations to investigate when and how AGN feedback affects its host galaxy. We isolate the impact of such feedback on stellar masses and large scale gas flows by comparing the evolution of a statistically representative sample of individual objects, identifying matching galaxies in two simulations, HORIZON-AGN (H-AGN) and HORIZON-noAGN (H-noAGN). As their names indicate, these simulations are identical in all aspects except one is run with and the other without AGN feedback. Following the evolution of twinned galaxies from redshift \(z=5 \) down to \(z=0 \) allows us to determine the epoch of quenching, and identify ensuing changes in the stellar masses of affected galaxies. 

The paper is structured as follows: Section \ref{sec:simulation} briefly introduces the HORIZON simulation suite and Section \ref{sec:twinning} explains the procedure used to identify pairs of corresponding objects across both simulations. Section \ref{sec:stellar_quenching} presents the effect of AGN feedback on galaxy stellar masses throughout cosmic time and Section \ref{sec:quenching_reasons} determines the causes for the measured quenching by studying the evolution of the black hole population, the gas content of halos and galaxies and gas inflow/outflow rates. Section \ref{sec:conclusions} summarises and discusses our results. 

%%%%%%%%%%%%%%%%%%%%%%%%%%%%%%%%%%%%%%%%
	
\section{The simulations}
\label{sec:simulation}
This paper presents a comparative analysis of two simulations: HORIZON-AGN (H-AGN) and HORIZON-noAGN (H-noAGN). Both simulations are run from identical initial conditions and share the same technical specifications and implementations of physics. The only difference is that HORIZON-AGN also includes a sub-grid modelling of SMBHs and the associated AGN feedback (see Section \ref{subsec:AGN_feedback}) whereas H-noAGN does not. More details can be found in \cite{Dubois2014}.

\subsection{Cosmology and initial conditions}
Both simulations were run with RAMSES \citep{Teyssier2002}, an adaptive mesh refinement code, using a second-order unsplit Godunov scheme to solve the Euler equations. A HLLC Riemann solver with a MinMod Total Variation Diminishing scheme was used to reconstruct interpolated variables. Initial conditions were produced using MP-GRAFIC \citep{Prunet2008} and both simulations were carried out until \( z=0.0 \).

The initial conditions setup is a standard $\Lambda$CDM cosmology consistent with the WMAP-7 data \citep{Komatsu2011}, with \(\Omega_{ \rm m}=0.272 \), dark energy density \(\Omega_{\rm \Lambda}=0.728 \), baryon density \(\Omega_{\rm b}=0.045 \), Hubble constant \(H_0=70.4 \: \mbox{kms}^{-1}\, \mbox{Mpc}^{-1} \), amplitude of the matter power spectrum \(\sigma_8=0.81 \), and power-law index of the primordial power spectrum \(n_s=0.967 \).

The simulated cube of \(L_{\rm box}=100 \: \mbox{Mpc/h}  \) on a side is initially refined uniformly down to physical \( \Delta x = 1  \) kpc, which requires a root grid with \(1024^3 \) cells at \( z \simeq 100  \). Extra refinement levels are continuously triggered using a quasi-Lagrangian criterion: a grid cell is split into eight whenever its Dark Matter (DM) or baryonic mass exceeds eight times the initial DM or baryonic mass respectively. To keep the size of the smallest cells approximately constant in physical units, cells on the most refined level are split into eight every time the expansion factor doubles, if they fulfil the refinement criteria. 

There are a total of \(1024^3 \) DM particles in each simulation, leading to a DM mass resolution of \(8 \times 10^7  M_\odot  \). All collisionless particles, i.e. DM and stars, are evolved using a multi-grid Poisson solver with a cloud-in-cell interpolation to assign particles to grid cells. 

\subsection{Cooling and heating}
The gas is allowed to cool down to \(10^4 \: \mbox{K} \) using H, He and atomic metal cooling, following \citet{Sutherland1993} and accounting for photon heating 
by a uniform UV background \citep{Haardt1996} from \(z_{\rm reion}=10 \) onwards. The ratio between elements is assumed to be solar in these cooling/heating calculations.  
The gas follows a mono-atomic equation of state, with adiabatic index \( \gamma = 5/3  \).  

\subsection{Stars and Supernovae}
Star formation is modelled using a Kennicutt--Schmidt law \( \dot{\rho}_* = \epsilon_* \rho / t_{\rm ff} \), where \( \dot{\rho}_* \) is the star formation rate density, \( \epsilon_*=0.02  \) the (constant) star formation efficiency \citep{Kennicutt1998,Krumholz2007}, $\rho$ the gas density and  \(t_{\rm ff}  \) the local free fall time of the gas. Stars form when the gas number density exceeds \(\rm \rho_0 /(\mu m_H) = 0.1\: {H/cm}^{3}  \) where $m_H$ is the mass of a hydrogen atom and $\mu$ the mean molecular weight, and star particles are generated according to a Poisson random process \citep{Rasera2006,Dubois2008} with a stellar mass resolution of \( M_* \simeq 2 \times 10^6 M_\odot  \), kept constant throughout the simulation. To avoid numerical fragmentation, and mimic the effect of stellar heating by young stars, a polytropic equation of state, \( T = T_0(\rho/\rho_0)^{\kappa-1} \), is used for gas above the star formation density threshold, with \( \kappa=4/3 \).

For stellar feedback, a Salpeter initial mass function (IMF) \citep{Salpeter1955} is assumed, with low and high mass cutoffs of \( 0.1 \) M\(_\odot  \) and \( 100  \) M\(_\odot  \) respectively. In an effort to account for stellar feedback as comprehensively as possible, the (sub-grid) model implemented in this work includes stellar winds, Type II and Type Ia supernovae.  Mechanical feedback energies from Type II supernovae and stellar winds are computed using STARBURST99 \citep{Leitherer2010,Leitherer1999}. Specifically, we use a Padova model \citep{Girardi2000} with thermally pulsating asymptotic branch stars \citep{Vassiliadis1993}, and stellar winds are calculated as in \citet{Leitherer1992}. The frequency of Type I SN is estimated from \citet{Matteucci1986}, assuming a binary fraction of 5\%.  To reduce computational costs, stellar feedback is modelled as a source of kinetic energy during the first 50 Myr of the lifetime of star particles, and as a heat source after that. On top of energy, mass and metals injected into the interstellar medium (ISM) by stellar feedback, we also keep track a variety of chemical elements (O, Fe, C, N, Mg, Si) synthesised in stars, with stellar yields estimated according to the W7 model of \citet{Nomoto2007}. More detailed discussions of the stellar feedback model used can be found in \citet{Dubois2008,Kimm2015,Rosdahl2017}.

\subsection{SMBH formation and accretion}
In H-AGN, black holes are seeded with an initial mass of \( 10^5  \) M\(_\odot  \) in dense, star-forming regions, i.e. when a gas cell exceeds \( \rho > \rho_0  \) and is Jeans unstable, provided such regions are located more than 50 kpc away from a pre-existing black hole \citep{Dubois2010}. These black holes subsequently accrete gas at the Bondi-Hoyle-Lyttleton rate:
\begin{equation}
	\label{eq:bondi_rate}
	 \dot{M}_{\rm BH}=\frac{4 \pi \alpha G^2 {M_{\rm BH}}^2 \bar{\rho} }{(\bar{c_{\rm s}}^2 + \bar{u}^2)^{3/2}},
 \end{equation}
where \(M_{\rm BH} \) is the black hole mass, \(\bar{\rho} \) is the average gas density, \(\bar{c_{\rm s}} \) is the average sound speed, and \(\bar{u} \) is the average gas velocity relative to the BH. We emphasize that this model is based on rather crude assumptions about the hydrodynamical processes undergone by gas surrounding an {\em extremely small} accretor. In particular, it does not take possible density or velocity gradients on the scale of the (unresolved) accretion radius into account, nor does it provide any description of important hydrodynamical instabilities which develop on yet 
smaller scales \citep[see e.g.][]{Foglizzo2005}. To somewhat mitigate resolution effects that make it difficult to capture cold, dense regions of the ISM, a boost factor $\alpha$ is used, following \citet{Booth2009} and \citet{Dubois2012}:
 \begin{equation} 
 	\label{eq:alpha}
	\alpha = 
 		\begin{cases}
			(\rho / \rho_0)^2 & \quad \text{if }  \rho >\rho_0 \\
			1 & \quad \text{otherwise}
		\end{cases}
\end{equation}
Accretion is capped at the Eddington rate \begin{equation}
\dot{M}_{\rm Edd}= 4 \pi G M_{\rm BH} m_{\rm p} / (\epsilon_{\rm r} \sigma_{\rm T} c) , 
\label{eq:eddington_rate}
\end{equation}
where \( \sigma_{\rm T} \) is the Thompson cross section, \(m_{ \rm p} \) is the proton mass, and \(c \) is the speed of light. A standard radiative efficiency, typical of a Shakura-Sunyaev accretion disc around the BHs, of \( \epsilon_{\rm r}=0.1 \) is assumed \citep{Shakura1973}. 

BHs are also allowed to merge with one another when they are closer than $4kpc$, and their relative velocity is smaller than the escape velocity of the binary. 

 \subsection{AGN feedback}
 \label{subsec:AGN_feedback}
Two modes of AGN feedback are implemented in H-AGN, depending on the instantaneous accretion rate of the SMBH: the so-called radio and quasar modes.

At high accretion rates, i.e. for Eddington ratios \(\chi= \dot{M}_{\rm BH}/\dot{M}_{\rm Edd} > 0.01 \), the quasar mode deposits thermal energy isotropically into a sphere of radius \( \Delta  \)x centred on the BH. This energy is deposited with an efficiency of $\epsilon_{\rm f} = 0.15 $ at a rate of 
\begin{equation}
	\label{eq:feedback_energy}
	\dot{E}_{\rm AGN}=\epsilon_{\rm f} \epsilon_{\rm r} \dot{M}_{\rm BH} c^2 .
\end{equation}

The radio mode takes over at low accretion rates, \(\chi= \dot{M}_{\rm BH}/\dot{M}_{\rm Edd} \leq 0.01  \), and deposits kinetic energy into bipolar outflows with jet/wind velocities of \( 10^4 \: \text{km/s} \), along an axis aligned with the angular momentum of the accreted material, following the model of \cite{Omma2003}.  The total rate of energy deposited is given by the previous equation for \( \dot{E}_{\rm AGN}  \), albeit using a higher efficiency of \( \epsilon_{\rm f}=1 \) (see \citet{Dubois2010} for detail). 

The radiative efficiencies of the two AGN feedback modes were chosen to reproduce the scaling relations between BH mass and galactic properties in the local universe, \(\rm M_{SMBH} - M_{*} \) and \(\rm M_{SMBH}-\sigma_{*} \) \citep{Dubois2012}. More generally, we refer the reader interested in the details of how accretion onto SMBHs, and subsequent feedback injection, depend on numerical resolution and 
sub-grid model parameter choices to \cite{Dubois2012}. Note that the two radiation efficiency parameters previously mentioned are the only ones which are tuned in the HORIZON simulations, in the sense that the 
other parameters (associated with the sub-grid models of star formation and stellar feedback) were not allowed to vary in order to obtain a better match to bulk galaxy 
properties. For instance, even though our star formation efficiency choice ensures that galaxies will fall on the Kennicutt observational 
law by construction, it does not automatically guarantee that they will have the correct stellar/gas mass and/or size at any epoch. 

\subsection{Mass categories for galaxies}
\label{subsec:categories}

\begin{table}
\setlength{\tabcolsep}{5pt}
\centering
	\begin{tabular}{  c | c | c |c }
		\hline
		\multicolumn{4}{| c |}{\bf{Number of twins per mass category}} \\
		\hline
 		z & Small & Medium & Large \\ 
		\hline
		& $\rm M_*\leq10^9 M_\odot$ & $\rm10^9 M_\odot< M_*<10^{11} M_\odot$ & $\rm  10^{11} M_\odot \leq M_*$ \\
		\hline
 		\(0.1\) & 9,508 & 51,652 & 2,060  \\  
 		\(1\) & 13,393 & 60,547 & 773\\
 		\(3\) & 11,823 & 19,855 & 7\\
 		\(5\) & 2,949 & 1,626&  0 \\
		
	\end{tabular}
	\caption{Number of galaxy twins in each sub-sample, at the range of redshifts presented here.}
	\label{tab:mass_numbers}
\end{table}

To facilitate the presentation of our results, we split our sample of galaxy twins into three sub-samples, distinguished by the stellar mass of the H-AGN galaxy. We define {\em small} galaxies as twins with stellar masses \(\rm M_*^{H-AGN} < 10^9 M_\odot  \) in H-AGN, {\em medium} galaxies as those with \( \rm 10^9 M_\odot \leq M_*^{H-AGN} \leq  10^{11} M_\odot  \) and {\em large} galaxies as those with \( \rm M_*^{H-AGN} > 10^{11} M_\odot  \). See Table \ref{tab:mass_numbers} for the number of twins in each mass category.  As a visual guidance, these mass categories will be annotated by solid vertical lines on all relevant plots.

\subsection{Nomenclature}
\label{subsec:quenching} 
For the purpose of this paper, ``quenching'' refers to any reduction in galaxy star formation rates (SFR) when AGN feedback is included, compared to the case without it, not just a redshift dependent specific SFR threshold of \(\rm 0.3 / t_{Hubble} \), as defined in e.g. \citet{Franx2008}. The ``quenching mass ratio'' refers to the stellar mass ratio of galaxy twins between the cases with and without AGN feedback, i.e. \( \rm M_{*}^{H-AGN}/M_{*}^{H-noAGN}\).

%%%%%%%%%%%%%%%%%%%%%%%%%%%%%%%%

\section{Halo matching across simulations}
\label{sec:twinning}

\subsection{The twinning procedure}

\begin{figure}
	\centering
	\includegraphics[width=0.5\textwidth]{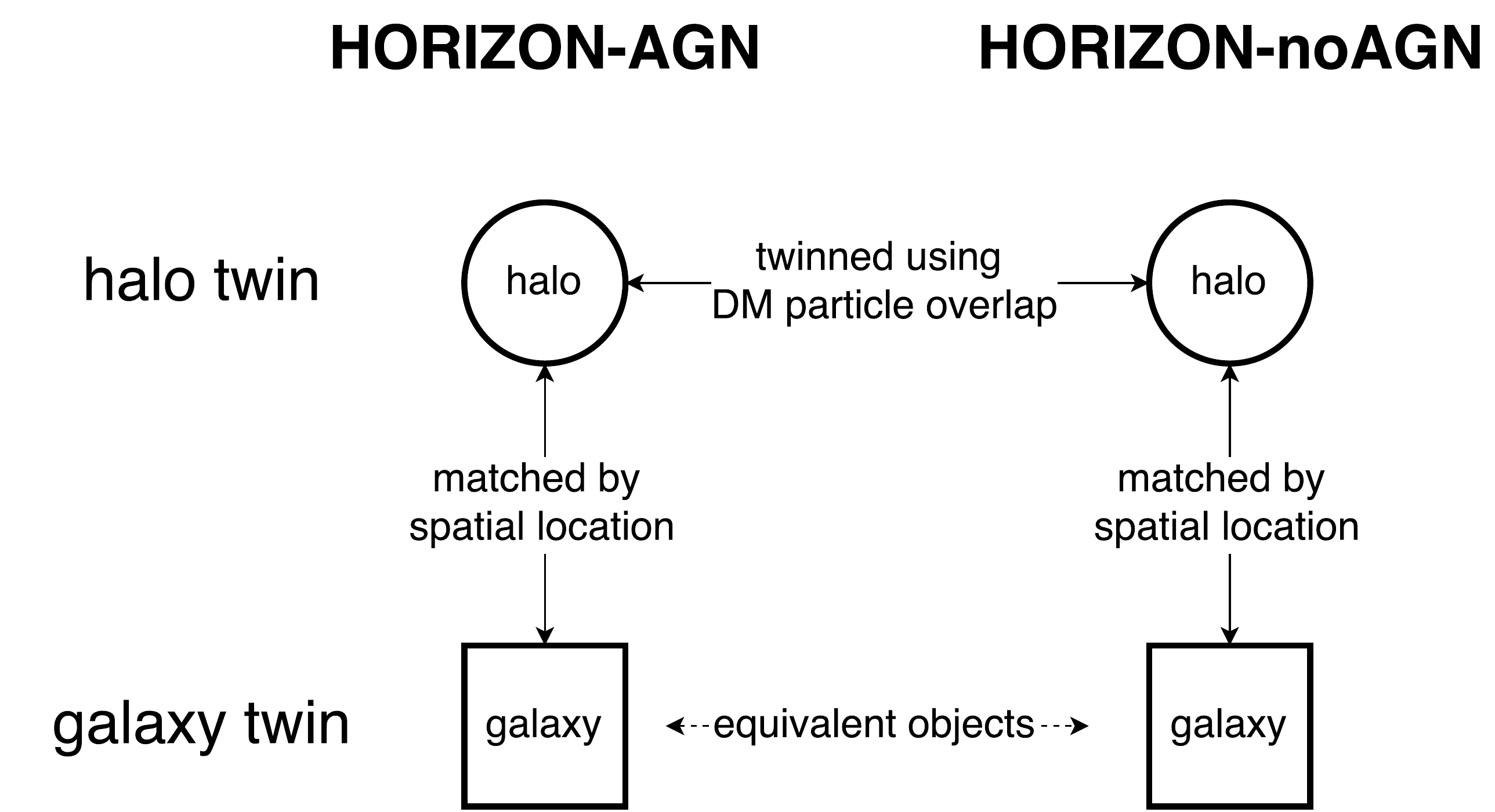}
	\caption{To twin galaxies across H-AGN and H-noAGN, DM (sub)halos are first identified in each simulation, and (sub)halos that show a 75\% or more overlap in the identities of their DM particles are twinned. Galaxies are then associated with a unique host (sub)halo in each simulation, based on how close they are to the centre of the (sub)halo (see text for detail). Galaxies associated with (sub)halo twins are considered galaxy twins.}
	\label{fig:twinning}
\end{figure}

As usual, the first step consists in detecting objects of interest (halos, subhalos and galaxies) in each simulation, using the {\sc adaptahop} (sub)halofinder \citep{Aubert2004,Tweed2009}.
Having two simulations based on identical initial conditions allows the identification of corresponding objects between the two simulations, a procedure here referred to as ``twinning" (see also e.g. \citealt{Geen2013}). A pair of corresponding objects is called a twin, and identifies two objects (one in each simulation) that have grown from the same overdensity in the initial conditions. These objects can either be (sub)halos, for (sub)halo twins, or galaxies, for galaxy twins. If all the algorithms implemented to describe physical processes were identical in both simulations, the twins would be identical except for minor differences introduced by the stochastic nature of the star formation algorithm. However, such seemingly innocuous differences would already prevent us from directly twinning galaxies. We therefore employ the more general method of \citet{Peirani2016} to perform this task, which is summarised in Fig. \ref{fig:twinning}. Only DM (sub)halos are twinned directly; to create galaxy twins, each galaxy is first associated to a host (sub)halo in the same simulation, before being twinned to the galaxy hosted by this (sub)halo's twin in the other simulation. 

More specifically, as both simulations start from identical initial conditions, with uniquely identified DM particles in identical positions, we can identify which of these particles cluster to form gravitationally bound (sub)halos as the runs proceed. (Sub)halos that grew from the same initial overdensities in both simulations should contain a large fraction of DM particles with identical identities at any time. In practice, for two DM (sub)halos to be twinned, we require that at least 75\% of the DM particles present in a (sub)halo in H-AGN are also present in the H-noAGN (sub)halo. Note that in some cases, this choice will lead to a single (sub)halo in H-noAGN being associated with several (sub)halos in H-AGN, as (sub)halos mergers lead to the formation of (sub)subhalos which are not necessarily disrupted at the same time in both simulations. In these cases, the object with the most similar mass is chosen as the twin (sub)halo, and the other matches are discarded.

Star particles stochastically form over the course of a simulation and their identifiers therefore reflect the detailed star formation history of that precise simulation, so it is not possible to identify galaxy twins directly through their star particle identities as is done for DM (sub)halos. Instead, galaxies are considered to be twins if they are located within DM (sub)halo twins (see Fig. \ref{fig:twinning}). We therefore begin by assigning a host (sub)halo to each galaxy if its centre is located within a distance \( \rm R_{host}=0.05 \times R_{\rm vir} \) of the centre of the (sub)halo. The centres of these (sub)halos are computed using a shrinking sphere method \citep{Power2003}, and their precise location corresponds to the position of the most dense DM particle located in the final sphere identified with this method \citep{Peirani2016}. In case a (sub)halo contains more than one galaxy in its central region, we select the most massive one as being hosted by this (sub)halo and discard other matching objects. Note that proceeding in this way biases our results against so-called 'orphan' galaxies, i.e. galaxies whose host (sub)halo has been disrupted to the point that it falls below the particle detection threshold used by our halofinder. However, such orphans are quite rare (less than 1\% of our sample at any redshift) and almost exclusively belong to the category of small galaxies ($M_{*} < 10^9$ M$_\odot$), so our conclusions are unaffected by this bias. We also find that relying on DM (sub)halo host twinning to twin galaxies, rather than directly measuring the  galaxy orbital properties, is a more robust process, as orbital parameters are sensitive both to internal changes in galaxy properties (in particular stellar mass) and host (sub)halo density profiles, which can differ quite significantly between simulations with and without AGN feedback (see e.g. \cite{Peirani2016}).  

\subsection{Matched fractions}

\begin{table}
\setlength{\tabcolsep}{3pt}
\centering
	\begin{tabular}{  c | c | c |c | c}
	\hline
	\multicolumn{5}{| c |}{\bf{Number of (sub)halos}} \\
	\hline
 	z & \(\rm N^{H-AGN}\) & \(\rm N^{H-noAGN}\)  & \(\rm N^{H-noAGN}/{N^{H-AGN}}\) & \( \rm N_{twinned} \)\\ 
	\hline
 	\(0.1\) &  88,171 & 85141  & 0.97 &  75,544 \\  
 	\(1\) & 89,930 & 88,488  & 1.02 & 83,176\\
 	\(3\) & 53,363 & 53,551 & 1.00 & 53,357\\
 	\(5\) & 16,042 & 16,065  & 1.00 & 15,818 \\
	\end{tabular}
	
	\centering
	\begin{tabular}{ c | c | c |c | c}
	\hline
	\multicolumn{5}{| c |}{\bf{Number of galaxies}} \\
	\hline
 	z & \(\rm N^{H-AGN}\) & \(\rm N^{H-noAGN}\)  & \(\rm N^{H-noAGN}/N^{H-AGN}\) & \( \rm N_{twinned} \)\\ 
	\hline
 	\(0.1\) & 75,651 & 73,450  & 0.97 & 63,220\\  
 	\(1\) & 80,588 & 82,238 & 1.02 & 72,713\\
 	\(3\) & 34,128 &  47,428 & 1.39 &  31,685\\
 	\(5\) & 6,255 & 7,757  & 1.24 & 4,575\\
	\end{tabular}
	\caption{Number of galaxies, (sub)halos and twins identified in H-AGN and H-noAGN. The sample includes all (sub)halos resolved by at least 500 DM particles, and all galaxies resolved by at least 100 star particles, hosted by these (sub)halos.}
	\label{tab:object_numbers}
\end{table}

As we are interested in the most massive objects, the full sample considered here includes all (sub)halos resolved by at least 500 DM particles, and all galaxies hosted in these (sub)halos that contain at least 100 star particles. Note that this latter criterion, contrary to the one for DM (sub)halos does not correspond to a strict stellar mass threshold as star particles can have different masses (integer multiples of the minimal stellar mass). Across all redshifts, H-AGN and H-noAGN contain a comparable, albeit not identical number of both galaxies and (sub)halos (see Table \ref{tab:object_numbers}). For this reason, the following analysis considers the H-AGN sample as the reference to analyse the effectiveness of the twinning algorithm.

\begin{figure}
\centering
\includegraphics[width=0.5\textwidth]{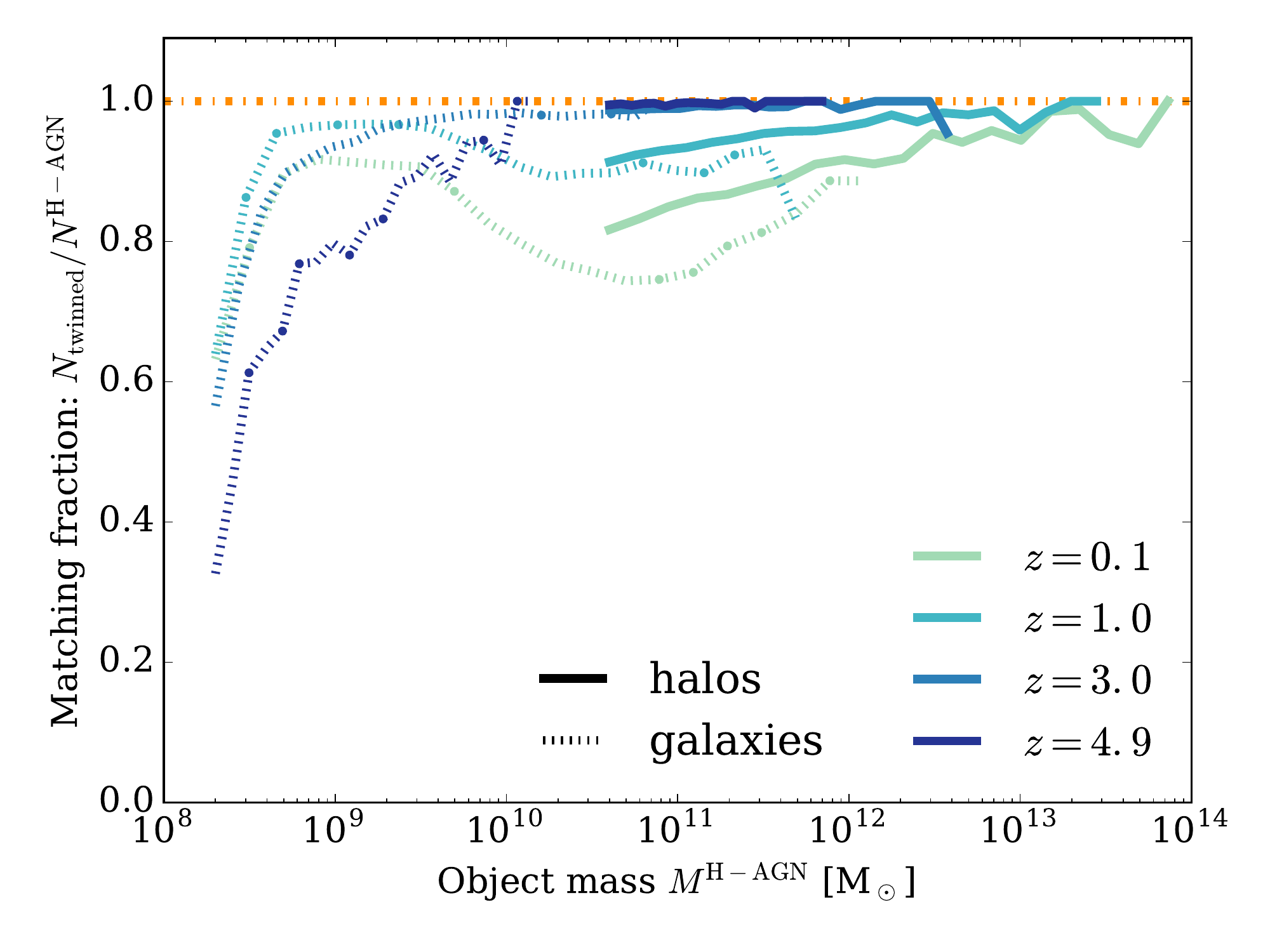}
\caption{Fractions of matched (sub)halos (solid lines) and galaxies (dashed lines) in the H-AGN and H-noAGN simulations. The dashed-dotted line represents the ideal case where all objects in H-AGN are uniquely associated with a corresponding object in H-noAGN. More than 86\% of (sub)halos are twinned, but galaxy twinning is less efficient, particularly at stellar masses below \(M_{*}^{H-AGN}= \rm10^9 M_\odot  \) for all redshifts but also, albeit less significantly, at redshift \(z \leq 1\) for galaxies with higher stellar masses. }  
\label{fig:match_numbers}
\end{figure}

Fig. \ref{fig:match_numbers} shows that at high redshift, over 98\% of (sub)halos present in H-AGN, corresponding to 15,818 out of 16,042 at redshift \(z=5 \), are twinned successfully, with an even distribution across all mass bins. At lower redshift, the fraction of matched (sub)halos decreases as the more and more different merger histories in the two simulations introduce larger discrepancies between individual objects. This fraction also decreases with (sub)halo mass, as (sub)halos with smaller particle numbers are more sensitive to these merger history changes. However, with 75,544 (sub)halos twinned at redshift \(z=0.1 \) out of a sample of 88,171, i.e. an average matched fraction of 86\%, the resulting sample remains statistically representative. 

The overall rates for matched galaxies are much lower than for (sub)halos, as they require three steps to establish the twin link (galaxy to host (sub)halo, host (sub)halo to host (sub)halo twin, host (sub)halo twin to galaxy twin, see Fig. \ref{fig:twinning}), with a number of objects dropping out of the sample at each step. Identifying a galaxy with its host (sub)halo can be challenging, especially in dense environments and at high redshift where interactions are more common. For example, increasing the size of the region within which a galaxy is associated with its host (sub)halo, \( \rm R_{host}  \), from  \( \rm R_{host}=0.05 \times R_{\rm vir} \) to  \( \rm R_{host}=0.10 \times R_{\rm vir} \), increases the fraction of matched galaxies at high redshift \citep[see e.g.][]{Chisari2017}. More specifically, for redshift \(z=3 \), the total number of galaxies with at least 100 star particles identified with a host (sub)halo with the same number of DM particles, would rise from 47,656 to 67,301, out of 76,887 galaxies in total, i.e. an increase in matched fraction from $61 \%$ to $87 \%$. Selecting amongst these galaxies those hosted in (sub)halos containing more than 500 DM particles (as we do in this work), further reduces the numbers to 34,128 (given in table \ref{tab:object_numbers}) and 48,354 galaxies respectively, out of 76,887. Whilst the number of galaxies excluded by the strict position and mass criteria we employ for twinning does represent a significant fraction of the sample, especially at high redshift, we have checked that relaxing them hardly alters the quantitative results presented in this work. This can be intuitively understood as the mass cuts chosen only eliminate low mass galaxies from the sample. Low mass galaxies are  both (i) the most numerous and (ii) virtually unaffected by AGN feedback, at any redshift. This can be seen in Fig. \ref{fig:GMF}, where the entire population (i.e. all 76,887 objects for H-AGN at $z=3$), as opposed to only twinned galaxies, is plotted. Similarly, the use of a stricter position criterion mostly affects low mass galaxies, as these have longer dynamical friction times and are more easily dislodged from the centre of their host (sub)halos during gravitational interactions. In short, the galaxy sample as defined in this section is statistically robust enough, at all redshifts and galaxy masses, for us to draw conclusions about the impact of AGN feedback in the H-AGN simulation, while allowing us to get rid of virtually all mismatch errors in the galaxy twinning process.

\subsection{The effect of AGN feedback on halos}
\label{sec:DM_halos}

\begin{figure*}
	\centering
	\includegraphics[width=\textwidth]{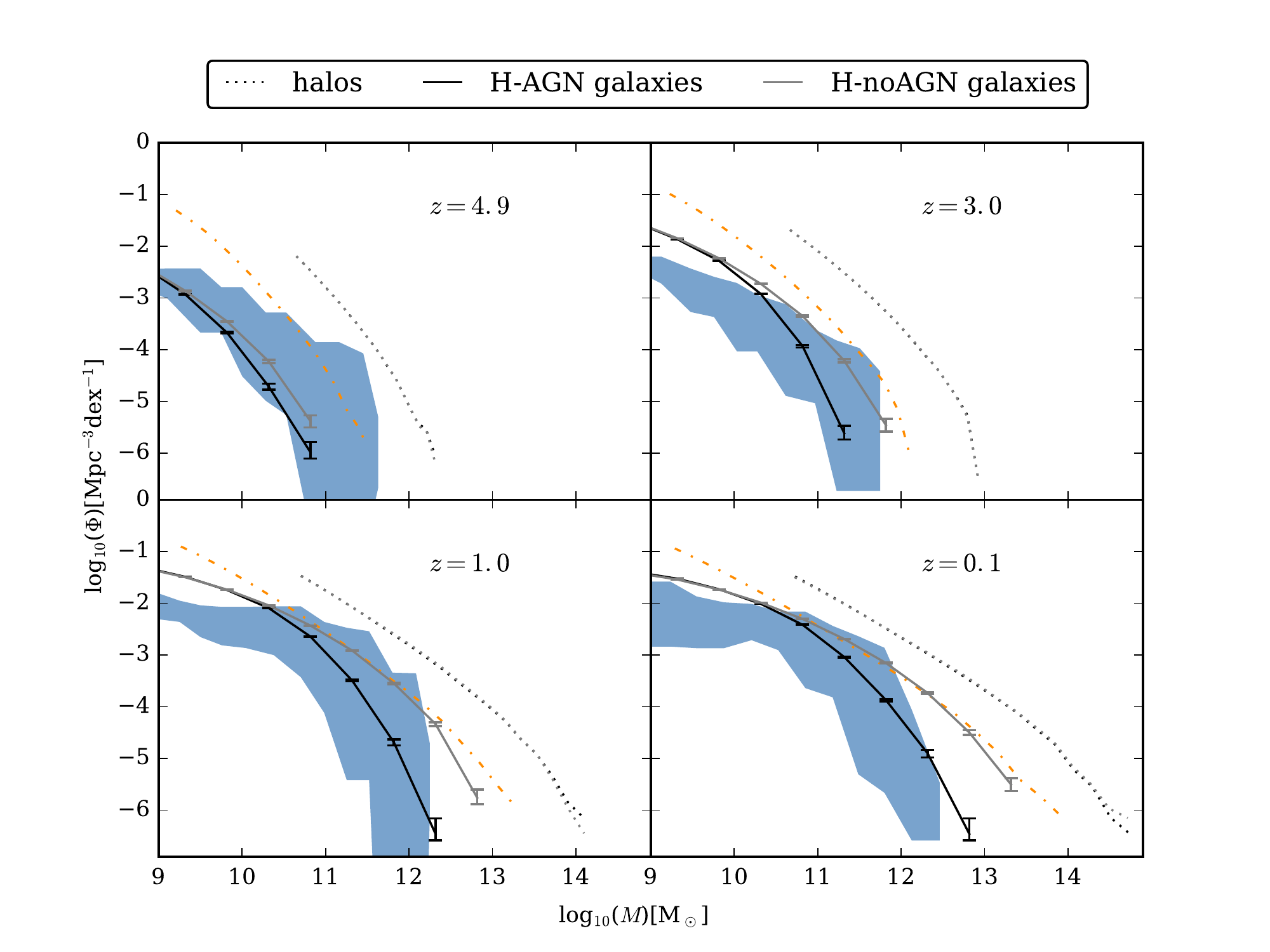}
	\caption{DM halo mass functions (HMF) (dotted) and corresponding galaxy stellar mass functions (GSMF) (solid) for H-AGN (black) and H-noAGN (light grey). We also plot combined observations from \protect\citet{Gonzalez2011,Bielby2012,Fontana2004,Moustakas2013,Tomczak2013,Bernardi2013,Song2015,Davidzon2017} as the shaded area, and the GSMF expected if all the baryons within (sub)halos (using a fraction of 16.5\% of the total mass) are converted into stars (dot-dashed line). A detailed comparison of individual observational datasets to the GSMF in H-AGN can be found in \citet{Kaviraj2016}. Errorbars on the simulated mass functions are Poisson, with those on the HMF not plotted as they are even smaller that the GSMF ones in the mass range where they overlap. Both HMF for H-AGN and H-noAGN are plotted, but are so similar as to be indistinguishable. Note that this Figure shows mass functions for the {\em full} sample of objects in H-AGN and H-noAGN, as opposed to the sample of twinned objects (see text for definition). At redshifts $z \leq 3$, the GSMFs for H-AGN lie within the region supported by observations for galaxies above stellar masses  $ \rm M_* > 5 \times 10^{10} M_\odot$. H-AGN and H-noAGN overproduce the number of smaller galaxies by up to a factor of 3, partially due to a lack of magnitude cut applied to the GSMF from simulations at low stellar masses but predominantly because of the low efficiency of the stellar feedback model implemented (see text for detail).}
	\label{fig:GMF}
\end{figure*}

\begin{figure}
%\thisfloatpagestyle{empty}
\centering
\includegraphics[width=0.5\textwidth]{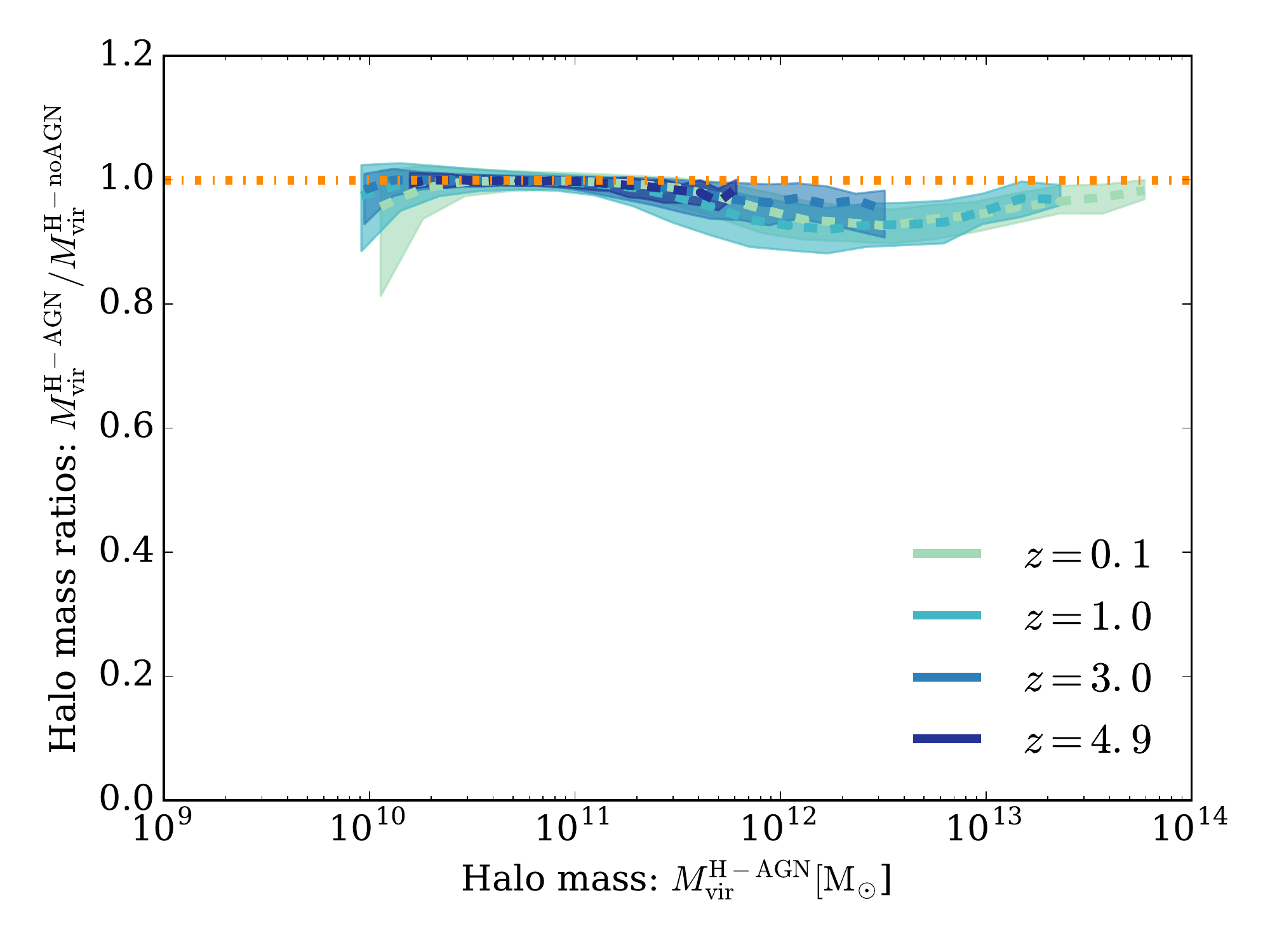}
\caption{Twinned (sub)halo DM mass ratios in H-AGN and H-noAGN at different redshifts, with quartile error regions
around the median (shaded region). Note that the median twin (sub)halo with \(\rm M_{vir}^{H-AGN} < 10^{11} M_\odot  \) has an identical DM mass with and without AGN feedback, whatever the redshift. Median intermediate mass (sub)halos  \(\rm 2 \times 10^{11} M_\odot < M_{vir}^{H-AGN} < 2 \times 10^{13} M_\odot  \) see their virial mass reduced by up to 5\% depending on mass and redshift.}
\label{fig:ratios_DM}
\end{figure}

Fig. \ref{fig:GMF} shows that no matter the redshift, the DM (sub)halo mass functions (HMF) for H-AGN and H-noAGN are so similar that they are indistinguishable on the plot. Directly comparing the DM masses of (sub)halo twins (See Fig. \ref{fig:ratios_DM}) shows that (sub)halos with masses below \( \rm M_{vir}^{H-AGN}< 10^{11} M_\odot  \) have identical masses in both simulations, at all redshifts. The small spread in masses is mainly caused by variations in shape, which the structure finding algorithm translates into a small variation in virial mass. At redshifts of  \(z=1 \) and below, (sub)halos with masses above \( \rm M_{vir}^{H-AGN} > 10^{11} M_\odot  \) can have a dark matter mass  up to 5\% percent lower in H-AGN. This is due to the fact that in the presence of AGN feedback, the baryon content of these massive (sub)halos is strongly reduced (see Section \ref{sec:quenching_reasons}), which translates into a reduced total (sub)halo mass, as the reduced gravitational pull slows down the cosmic inflow rate. As the (sub)halos in H-AGN systematically exhibit lower masses, it makes sense to require, as we do, that 75\% of DM particles from H-AGN be present in the H-noAGN (sub)halo, and not the reverse. A more detailed analysis as to how AGN feedback affects the inner structure of DM (sub)halos is carried out in \citet{Peirani2016}.  

%%%%%%%%%%%%%%%%%%%%%%%%%%%%

\section{AGN feedback \& stellar mass}
\label{sec:stellar_quenching}

\subsection{The galaxy stellar mass function}
\label{subsec:GSMF}
Comparing the galaxy population in H-AGN and H-noAGN at the various redshifts presented in this work shows that AGN feedback is instrumental in bringing the high mass end of the GSMF in agreement with observations\footnote{\label{note:GMF_obs} A more detailed comparison of the GSMF in H-AGN to individual observational datasets can be found in \citet{Kaviraj2016}. }. As Fig. \ref{fig:GMF} demonstrates, AGN feedback is able to suppress star formation in galaxies with masses \( \rm M_* \geq 5 \times 10^{10} M_\odot  \) at \(z = 3 \), allowing the simulation to match the number of galaxies at and above the knee of the GSMF. It is important to note that H-AGN was {\em not} tuned to reproduce this result. As previously mentioned, the only tuning done on global galaxy properties in the simulation involves the radiative efficiency of the AGN feedback modes, which were set to reproduce the local \(\rm M_{SMBH} - M_* \) and \( \rm M_{ SMBH} - \sigma_*  \) relations.

In the absence of AGN feedback, the GSMF in H-noAGN agrees well with predictions that the uniform baryon mass fraction of \(\rm \Omega_b / \Omega_m = 0.165 \) \citep[dotted-dashed lines,][]{Komatsu2011} is entirely converted into stars in the galaxy mass range \(\rm 5 \times 10^{10} M_\odot < M_* < 10^{12} M_\odot  \) by $z=1$. For galaxies with masses  \( \rm M_* \geq 5 \times 10^{12} M_\odot  \), a discrepancy between the GSMF in the absence of feedback (solid grey line) and expectation values from the cold dark matter model (dotted line) starts to appear because gas cooling times in host (sub)halos harbouring such massive galaxies become comparable to the Hubble time when the halos assemble, so not all the baryons enclosed have yet been able to cool and form stars. Cooling is further hampered by the fact that in the absence of AGN feedback, heavy elements do not get distributed effectively throughout the halo but remain close to the central galaxy. 

The simulations systematically overproduce the number of galaxies with masses below \( \rm M_* \leq 5 \times 10^{10}  M_{\odot} \) and $z \leq 3$. This is partly caused by the fact that observed mass functions are derived from magnitude-limited data, whereas the GSMF presented here for H-AGN and H-noAGN are raw stellar masses extracted from the simulation, with no completeness, surface brightness or luminosity cut applied. Indeed, comparing the GSMFs presented in Fig. \ref{fig:GMF} to those plotted in Fig.7 of \citet{Kaviraj2016}, which are based on the {\em same} simulation, H-AGN, but include a magnitude cut to match observations, one realises that the effective number of galaxies with masses
\( \rm M_* = 10^{9}  M_{\odot} \) is reduced by about 0.1-0.2 dex depending on redshift, whilst galaxies with masses \( \rm M_* \geq 5 \times 10^{10}  M_{\odot} \) are completely unaffected. This does somewhat flatten the simulated GSMFs at the faint end, bringing them in better agreement with the data. However, the remaining discrepancy of about 0.3 dex with the data for galaxies with stellar masses \( \rm M_* \leq 10^{10}  M_{\odot} \) can probably be attributed to the implementation of an insufficiently energetic stellar feedback model, coupled with numerical resolution effects \citep{Kaviraj2016}\footnote{ \label{note:SNandBH} Note that a stronger stellar feedback is likely to affect black hole masses as well \citep{Dubois2015,Habouzit2017}.}, although how efficient such a feedback can realistically be is still a matter of debate. Having said that, as AGN feedback, through the comparison of H-AGN and H-noAGN, is measured to have no effect on the low mass end of the GSMFs (as clearly visible in Fig. \ref{fig:GMF}), and the limitations discussed above are present in both simulations, they likely have a very limited impact on the work presented here. Still, the few absolute measurements presented here, such as outflow rates, have to be examined bearing in mind that stellar feedback is probably underestimated in the simulations, and thus that values derived for galaxies with stellar masses \(  \rm M_* \leq 5 \times 10^{10} M_\odot  \) are very likely too low.

Finally, both simulations systematically underproduce massive (\(  \rm M_* \geq 5 \times 10^{10} M_\odot  \)) galaxies at redshift $z \geq 5$ (Fig. \ref{fig:GMF} top left panel). 
Given that the HMF multiplied by the universal baryon fraction (dot-dashed curve on the figure) seems to describe the observed data fairly well for galaxies in this  mass range,
this suggests that our inability to resolve the progenitors of halos early enough leads to star formation being artificially postponed. Obviously, since the gas content 
of these halos is still correctly estimated, galaxies will eventually catch up: their star formation rate will be slightly higher than expected, as long as more gas is present.
However, at high redshift, galaxy star formation timescales cannot be considered small in comparison to the time elapsed since their host halo formed, 
so their stellar masses can be significantly underestimated. Note that this resolution effect has completely vanished, at least for massive galaxies, by $z=3$.  
Moreover, as this issue affects H-AGN and H-noAGN in the same way, it cancels out in the comparative analysis of the two simulations that we perform in this work.

\subsection{Quenching}

\begin{figure}
\centering
\includegraphics[width=0.5\textwidth]{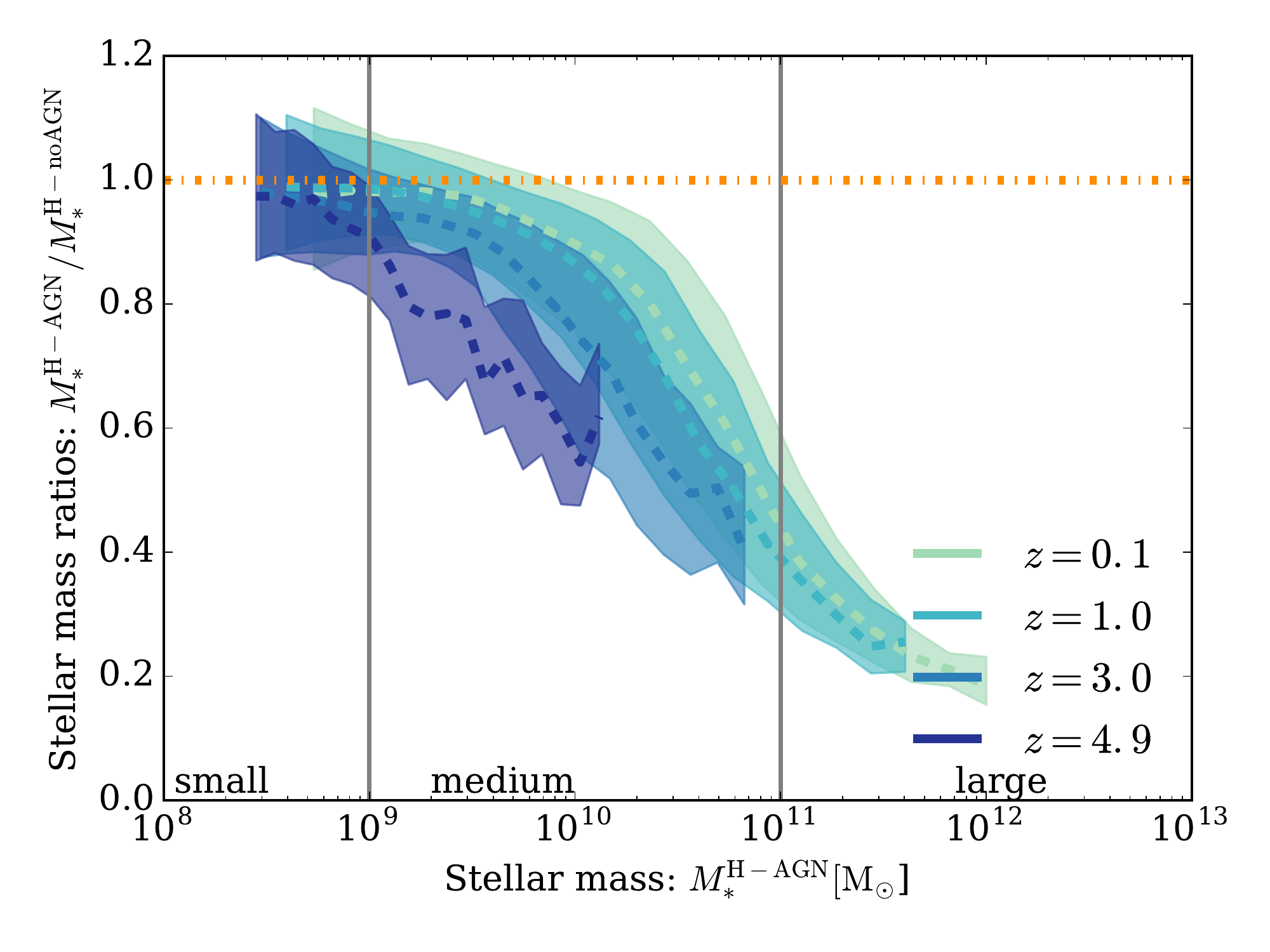}
\caption{Quenching mass ratios, i.e. stellar mass ratios of twinned galaxies in H-AGN and H-noAGN. The twins are mass binned by stellar mass in H-AGN, \( \rm M_*^{H-AGN}  \), with the median value (dashed lines) and quartile error ranges (shaded region) plotted for each bin. If AGN feedback had no effect, the quenching mass ratio would lie on the horizontal dotted line, which denotes twin galaxies having identical mass. Note the non-linear evolution of the quenching mass ratio with galaxy mass, with small galaxies (\(\rm M_*^{H-AGN} < 10^9 M_\odot  \)) barely affected, and large galaxies  (\( \rm M_*^{H-AGN} > 10^{11}M_\odot  \)) being most strongly quenched. This plot also shows a redshift evolution, with smaller galaxies more affected at higher redshift.}
\label{fig:ratios_mass_quartile}
\end{figure}

Instead of having to rely on statistical averages, such as those presented in the mass functions in Fig. \ref{fig:GMF}, the twinning procedure described in Section \ref{sec:twinning} allows for a direct comparison of the stellar masses of each individual galaxy, with and without AGN feedback. Fig. \ref{fig:ratios_mass_quartile} shows the results of this comparison for a range of redshifts.  As expected from the local GSMF in Fig. \ref{fig:GMF}, the most massive galaxies are the most strongly quenched at all redshifts (Fig. \ref{fig:ratios_mass_quartile}). However the amount of quenching does not vary linearly with galaxy mass, with the function tailing off for both strongly quenched large galaxies, and barely affected small galaxies. 

We also measure a redshift dependence in the maximum amount of quenching observed, ranging from 40\% for galaxies with stellar masses in H-AGN of  \(\rm M_*^{H-AGN} > 10^{10} M_\odot  \) at redshift \(z=5  \), up to a maximum of over 80\%  for the largest galaxies at redshift \(z=0.1  \), suggesting that galaxy quenching is a continuous process active throughout the merging history of galaxies. In general, the shape of the distribution is driven by small galaxies that show little influence of AGN feedback at any redshift, and a tailing off for massive galaxies. Large galaxies appear to converge to a constant quenching mass ratio \( \rm M_*^{H-AGN}/ M_*^{H-noAGN} = 0.2 \), as they grow from \(\rm M_* = 10^{11} M_\odot \) to \( \rm M_* = 10^{12} M_\odot  \) both in H-AGN and H-noAGN. This is not due to any fundamental change in the impact of AGN feedback for galaxies with stellar masses above \(\rm M_*>10^{11} M_\odot  \) but rather reflects the fact that, even in the absence of AGN feedback, the GSMF in H-noAGN steepens due to long cooling times for massive objects (see Fig. \ref{fig:GMF} and Section \ref{subsec:GSMF}) which lead to reduced star formation rates. Therefore, the constant quenching mass ratio for large galaxies is not driven by less effective AGN feedback, but rather by less effective cooling for galaxies in the absence of feedback. Note that this is not a selection effect either, as no mass ratio cut was applied to our galaxy sample.

Particularly noticeable for redshifts above \(z>3 \), the minimum mass to experience quenching decreases with redshift: a  typical \(\rm 10^9 M_\odot  \) galaxy at redshift \(z=5 \) already has its stellar mass quenched by 10\%, whereas a \(\rm 10^9 M_\odot  \) galaxy at redshift \(z=1  \) shows a median reduction in stellar mass of less than 1\%.  

\begin{figure}
	\centering
	\includegraphics[width=0.5\textwidth]{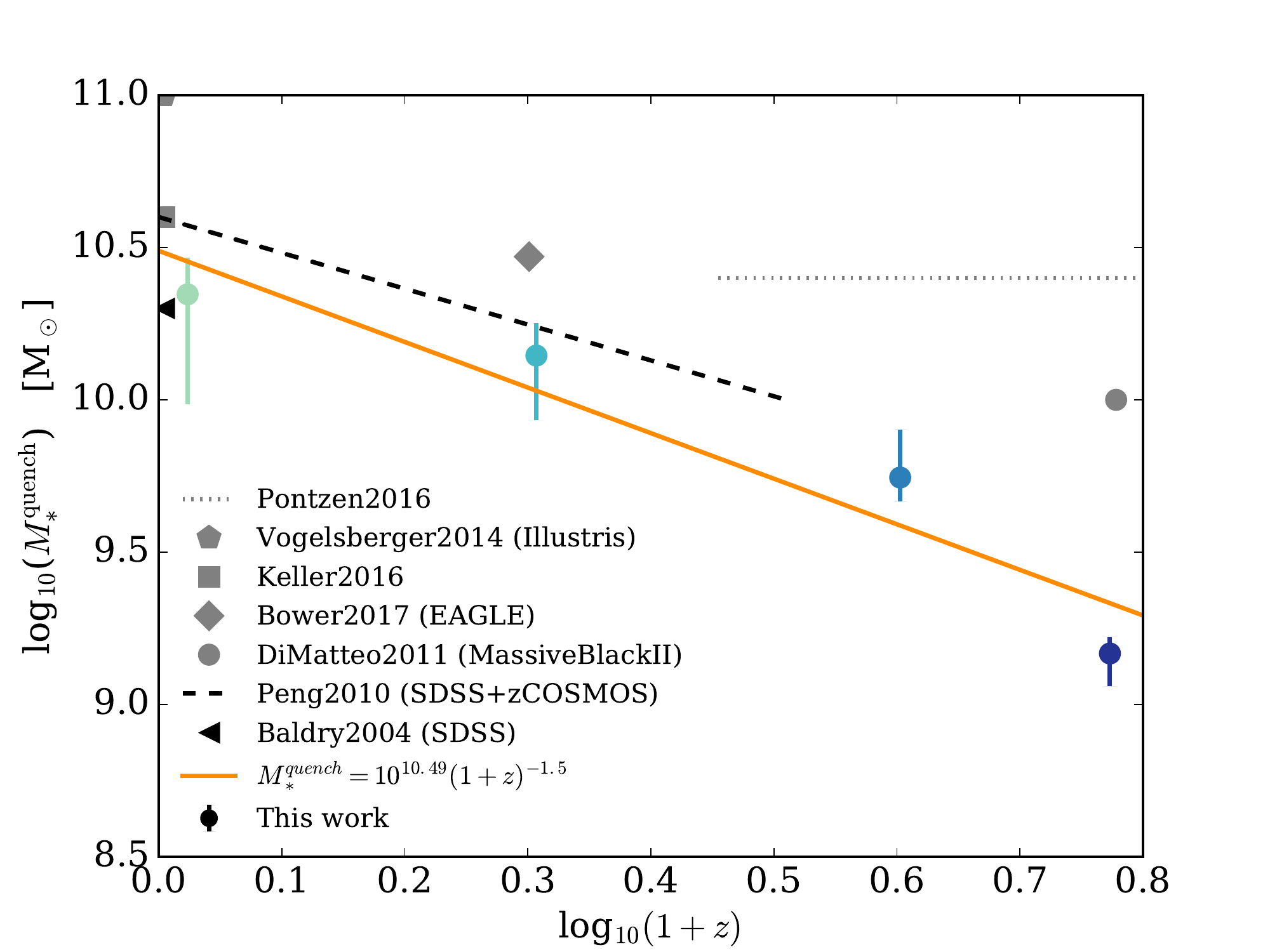}
	\caption{Transition galaxy stellar mass at which AGN feedback becomes important, measured where the quenching mass ratio equals $0.85 \pm 0.05 $, as well as observational data from \citet{Peng2010,Baldry2006} (black), and results from from a range of simulations \citet{Bower2017,DiMatteo2011,Keller2016,Pontzen2016} (grey). The values for \citet{DiMatteo2011} and \citet{Pontzen2016} are converted from the halo mass quoted in the paper, to stellar mass plotted here, using the $M_{vir}-M_*$ relation derived by \citet{Moster2013}.}
	\label{fig:transition_mass}
\end{figure}

A transition mass between star forming and quenched galaxies is somewhat difficult to define in this context, as the quenching mass ratio is a cumulative measure, not an instantaneous one such as the star formation rate or the stellar mass growth timescale often used to separate star forming and quiescent populations. However, based on Fig. \ref{fig:ratios_mass_quartile}, it seems natural to define the mass at which quenching due to AGN feedback becomes important to be the point where the quenching mass ratio equals $0.85 \pm 0.05$, as it corresponds to the location of the sharp break in the quenching mass ratio versus galaxy stellar mass relation. As can be seen in Figure \ref{fig:transition_mass}, this quantity shows a redshift evolution, decreasing from  $\log(M^{quench}_*/[M_\odot]) = 
10.35^{0.12}_{0.36} $ at $z=0.1$ to $\log(M^{quench}_*/[M_\odot]) = 9.17^{+0.05}_{-0.11} $ at $z=5$. This result is in good agreement with the value of $10.3$ found for a sample of SDSS galaxies by \citep{Baldry2004}. At higher redshift, $z \approx 1$, we find good agreement with the transition mass quoted by \citet{Peng2010}, who used a large sample of SDSS and zCOSMOS galaxies to study the stellar mass at which `` mass quenching'', which includes the impact of AGN, becomes important.  Note that our redshift evolution also suggests that the trend they observe can be extrapolated to $z=2$ at least. However, mostly driven by results above $z>4$, we find that our best fit power law for the redshift evolution,
\begin{equation} 
M_*^{quench} (z) = 10^{10.49} (1+z)^{-1.50}.
\end{equation} 
has a steeper slope than that reported in \citet{Peng2010}. Comparing to other simulations \citep{Bower2017,DiMatteo2011,Keller2016,Pontzen2016}, we consistently find a lower value. This discrepancy is partially due to how the transition mass is defined in each work. The value in \citet{Pontzen2016}, for example, is defined to be the value at which only AGN feedback quenches the galaxy. It is not surprising that this is higher than our definition of $M^{quench}_*$, which is the transition mass at which AGN feedback begins to be dominant, but stellar feedback might still play a role. 
We take the fact that we consistently measure lower transition masses to be evidence that quenching is a long term process, whose effect accumulates over the evolution history of a galaxy. A galaxy whose star formation rate is reduced by $20 \%$ would not sufficiently change colour to be counted among the ``red and dead'' population, but could over time show a noticeably reduced quenching mass ratio. The transition mass based on the cumulative star formation history is therefore necessarily lower than that based on a more instantaneous measure, such as the stellar mass growth timescale \citep{Bower2017} or the star formation rate \citep{Pontzen2016}.

\begin{figure}
	\centering
	\includegraphics[width=0.5\textwidth]{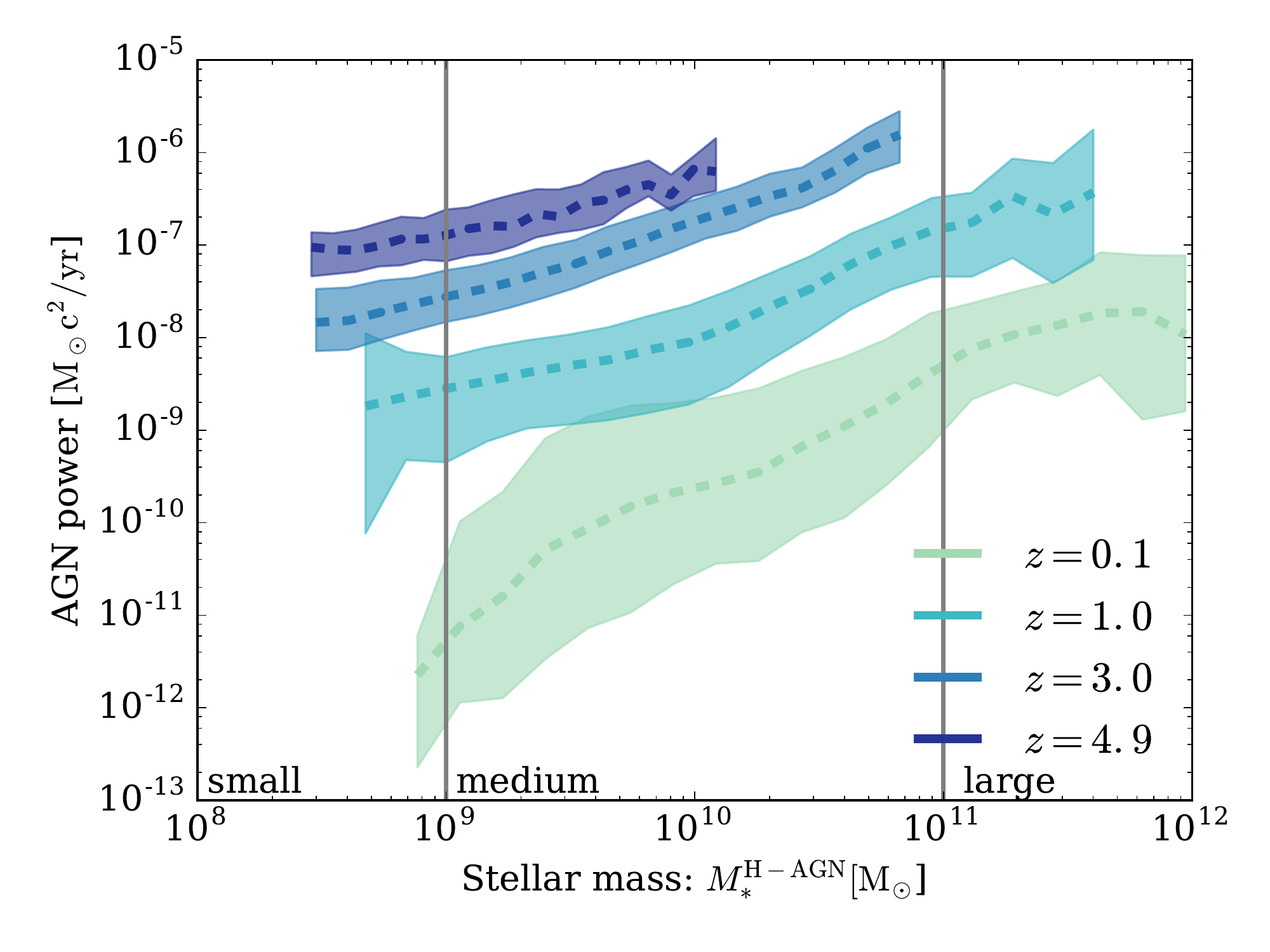}
	\caption{Median AGN power \(\rm \dot E_{AGN} \) (see equation \ref{eq:feedback_energy}) for each galaxy stellar mass bin in H-AGN, with the shaded region showing the samples quartile error ranges. At higher redshift, a galaxy of the same mass is subject to stronger feedback.}
	\label{fig:AGNpower_mass}
\end{figure}

As the evolution in the quenching mass ratio (and the minimum mass of a galaxy affected) is in our case case purely driven by AGN feedback, we expect the impact of AGN feedback to evolve with time. The evolution of the instantaneous AGN power, plotted in Fig. \ref{fig:AGNpower_mass} confirms this conjecture. Galaxies of the same stellar mass are subject to different amounts of feedback at different points in cosmic history. The median AGN power for a galaxy of a given stellar mass decreases strongly with redshift, with a galaxy with $\rm M^{H-AGN}_* = 10^9 M_\odot$ at $z=5$ subject to a feedback three orders of magnitude stronger than a galaxy of equivalent stellar mass at $z=0$. 

As the sound speed of the gas, and its relative velocity with respect to the SMBH, do not vary systematically with redshift, the feedback power of an AGN, calculated according to equation \ref{eq:feedback_energy} is mainly a function of local gas density in the vicinity of the BH, the Eddington ratio which determines the radiative efficiency of each mode of feedback and BH mass squared (the case of BHs accreting at the Eddington limit is rare and very short lived in the simulations, see e.g. \cite{Dubois2012} or section \ref{subsec:BHevolution} below). Therefore, the decreasing importance of AGN feedback in the evolution of small galaxies could be due to (i) the decreasing gas fractions associated with galaxy evolution, (ii) a shift in AGN feedback mode from quasar (high redshift) to radio dominated (low redshift), or (iii) an evolving black hole population (different \( \rm M_{\rm SMBH}  \) vs \( \rm M_* \) relation). We examine each of these three options in turn in the following section. 

\subsection{The coevolution of SMBHs and their hosts}
\label{subsec:BHevolution}

\begin{figure}
\centering
\includegraphics[width=0.5\textwidth]{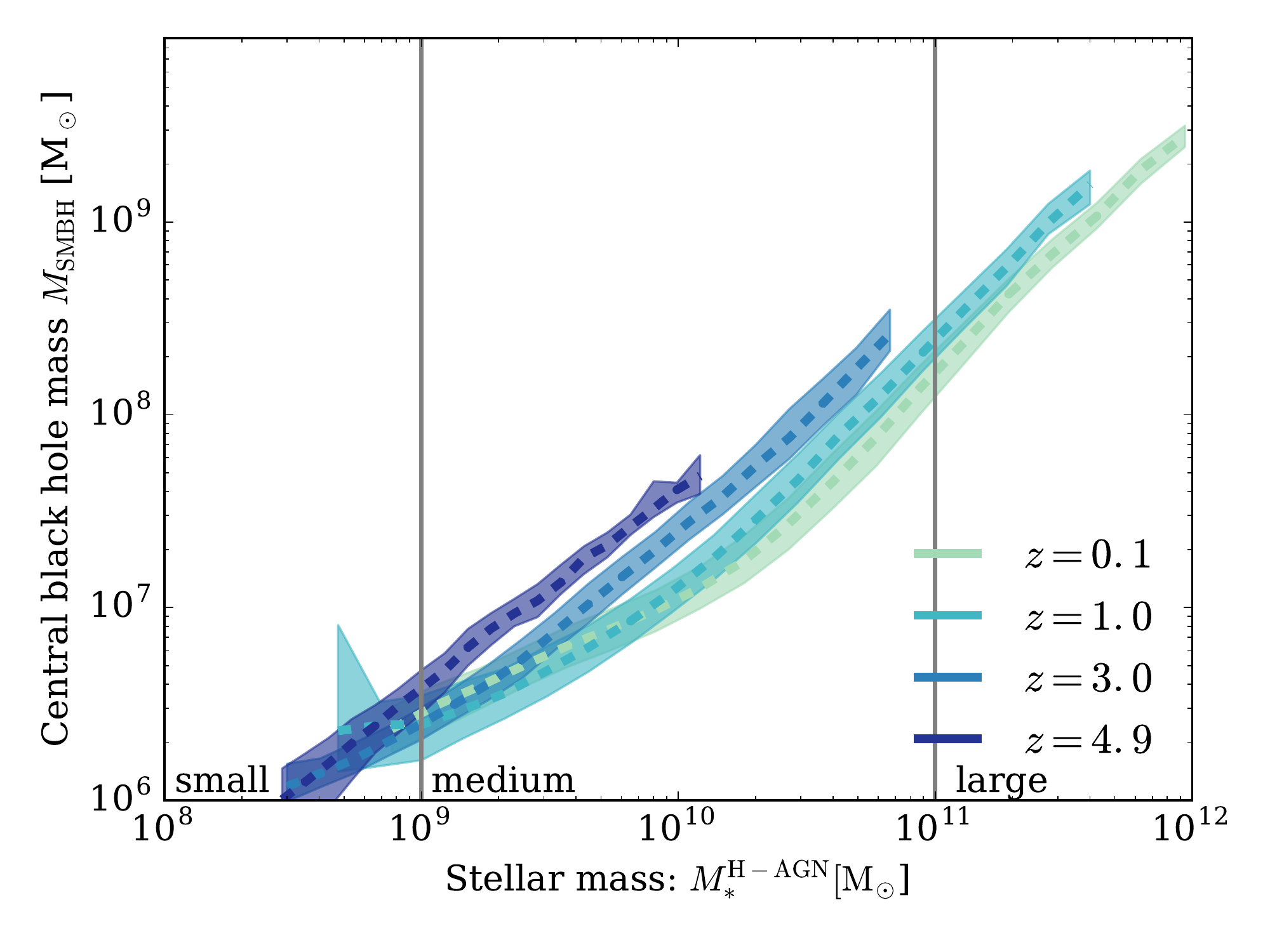}
\caption{Median black hole mass (dashed lines) evolution as a function of galaxy stellar mass in H-AGN, as well as quartile ranges for the BH mass distribution (shaded region). The plot shows a clear redshift evolution, with smaller galaxies hosting larger black holes at higher redshift \protect\textsuperscript{\ref{note:BHredshift}}. Note that small galaxies with stellar masses below \(\rm M_*^{H-AGN} < 10^{9} M_\odot  \) are to be treated with some caution, as they are quite sensitive to resolution effects. }
\label{fig:BHmass_mass}
\end{figure}

The total AGN power is a function of the accretion rate onto the central SMBH, which in turn, in the Bondi regime, depends on the SMBH mass squared\footnote{Or only of mass when accreting at the Eddington limit, but these episodes are rare and short lived in simulations with AGN feedback as we show later in this section; see equation \ref{eq:feedback_energy}, using equation \ref{eq:bondi_rate} or equation \ref{eq:eddington_rate}}. Fig. \ref{fig:BHmass_mass} shows that the median mass of the central black hole undergoes a redshift evolution between \(z=0\) and \(z=5\), with galaxies of a given stellar mass hosting a more massive BH at higher redshift\footnote{\label{note:BHredshift} The results presented here are consistent with fig. 10 of \citet{Volonteri2016}, which reports no redshift evolution in the \(\rm M_* -  M_{SMBH}  \) relation. This is because the redshift evolution we see is mainly driven by low mass black holes at low redshift, a sample excluded by these authors who apply a cut in host halo mass of \(\rm M_{halo}= 8 \times 10^{10} M_\odot  \). By comparison, the sample analysed here includes all black holes identified within a host galaxy in halos with \(\rm M_{halo}> 4 \times 10^{10} M_\odot  \). The second difference between these two pieces of work concerns the statistical analysis chosen: while \citet{Volonteri2016} employ a linear fit as used in observational studies, we present median black hole masses and thus allow for a non-linear correlation between black hole and galaxy stellar masses.}. For example, a galaxy with \(\rm M_*^{H-AGN}=10^{10} M_\odot \) at redshift $z=0.1$ typically hosts a SMBH with \(\rm M_{SMBH}=\rm 1.1 \times 10^{7} M_\odot \), whereas a galaxy with the same stellar mass at \(z=5\) hosts a SMBH with a median mass of \(\rm M_{\rm SMBH} = 4.2 \times 10^{7} M_\odot \). A similar evolutionary trend is reported observationally \citep{Merloni2010,Decarli2010} and in other large scale cosmological simulations \citep{Khandai2015,Sijacki2015}. A detailed discussion of the shape of the $\rm M_{SMBH}-M_*$ relation can be found in \citet{Volonteri2016}.

\begin{figure}
\centering
\includegraphics[width=0.5\textwidth]{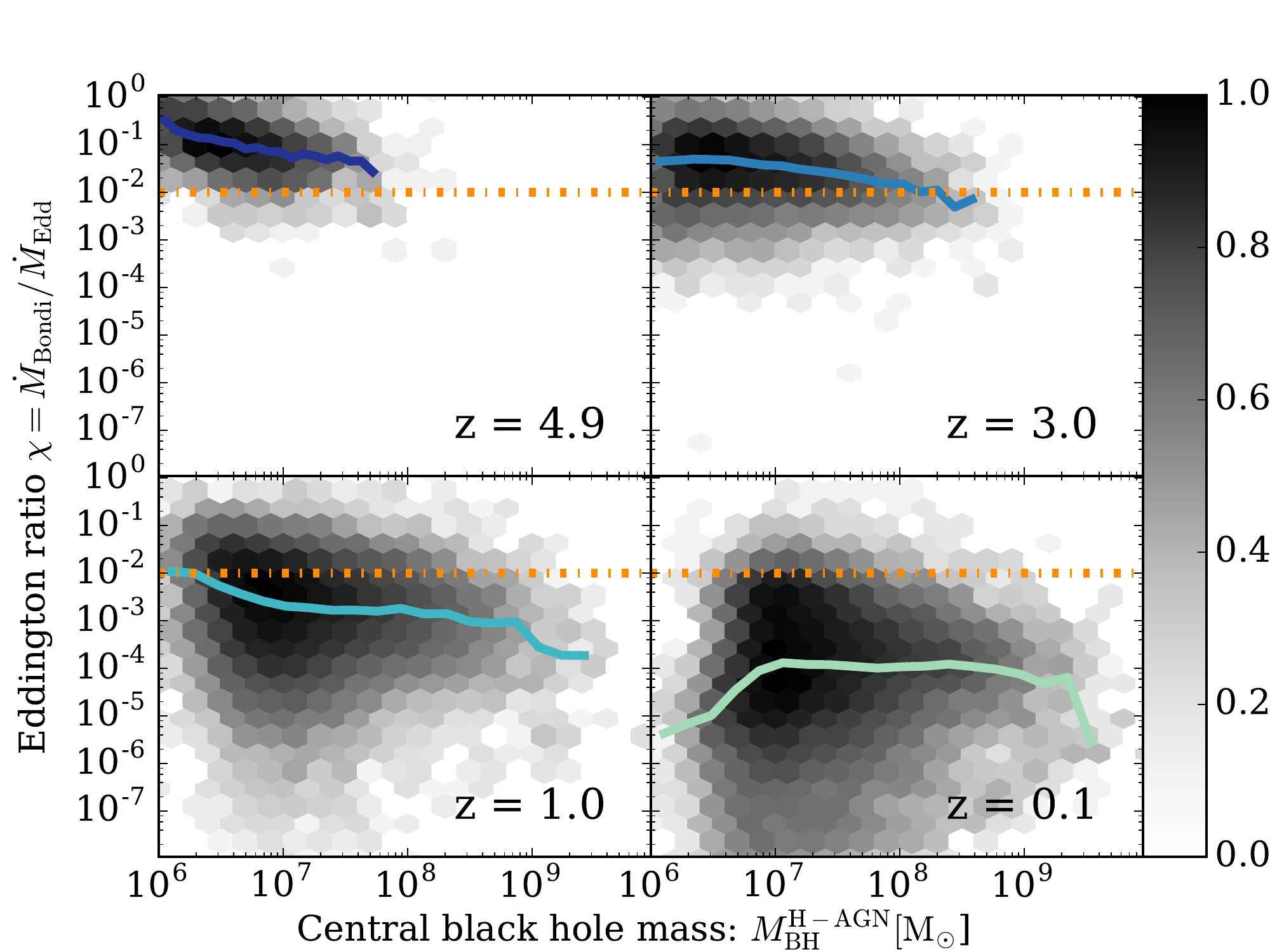}
\caption{Evolution of AGN feedback mode for the whole sample of SMBHs at different redshifts, with objects above the dotted line in quasar mode and objects below in radio mode. The solid lines denote the median Eddington ratio. There is a clear redshift evolution of the feedback mode, with 95.7\% of SMBHs in quasar mode at \(z=5 \), in comparison to only 2.3 \% at $z=0.1$. The colourbar in each panel is normalised to the bin with the maximum number of objects at the given redshift.}
\label{fig:scatter_eddington}
\end{figure}

Looking at the evolution of AGN feedback mode with redshift, as determined by the Eddington ratio $\chi$, Fig. \ref{fig:scatter_eddington} reveals that the SMBH population transitions from quasar mode to radio mode between redshifts \(z=3 \) and \(z=1 \), as $\chi$ falls below $0.01$. At high redshift, the vast majority of AGNs are found in quasar mode, namely 95.7\% of the sample at \(z=5 \) and 85.5\% at \(z=3 \). At lower redshift, the population is overwhelmingly in radio mode across all mass bins, with only 19.1\% and 2.3\% found in quasar mode at \(z=1 \) and \(z=0.1 \) respectively. 

As the Eddington ratio is a measure of how efficiently a black hole of a given mass is accreting, the high Eddington ratios at redshift $z>3$ explain the evolution in the $\rm M_{\rm  SMBH}-M_*$ relation in Fig. \ref{fig:BHmass_mass}. Indeed, whilst a black hole with \(\rm M_{\rm SMBH}=10^7 M_\odot  \) at redshift \(z=5 \) accretes with a mean Eddington ratio of \(\chi = 6.27 \times 10^{-2}  \), a SMBH with the same mass at redshift \(z=0.1 \) accretes at only \(\chi = 1.2 \times 10^{-4}  \) Eddington. This means that the latter grows about two orders of magnitude more slowly than the former. 

\begin{figure}
 \centering
 \includegraphics[width=0.5\textwidth]{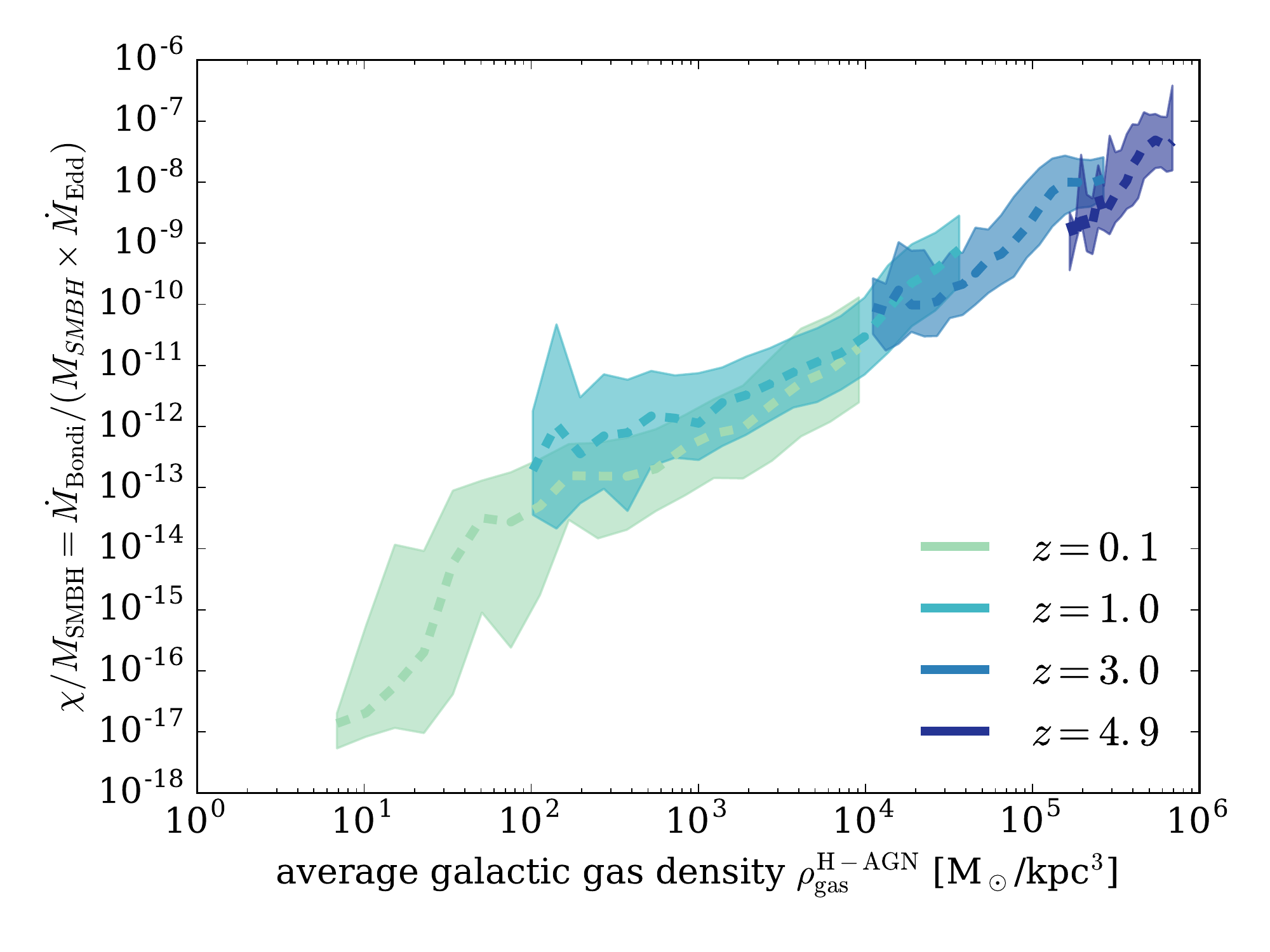}
 \caption{Specific accretion rate onto the central SMBH in Eddington units vs average gas density of the host galaxy for various redshifts. The measured decrease in average specific accretion rate for SMBHs of any mass, as redshift decreases, is driven by a decreasing gas density in the host galaxy.}
 \label{fig:eddington_density}
\end{figure}

Fig. \ref{fig:scatter_eddington} also clearly shows that except for a few outlying objects, the bulk of the population is not accreting in an Eddington limited fashion at $z \leq 5$. This evolution in the median Eddington ratio reflects an underlying evolution in the gas density of galaxies, as can be seen in Fig. \ref{fig:eddington_density} where we have divided $\chi$ by the SMBH mass to calculate the specific accretion rate of these SMBHs in Eddington units, i.e. we have removed the dependence on BH mass to be able to intercompare BHs across the whole mass range. The story which emerges from this plot is that since high redshift galaxies are more gas rich, they fuel their central black holes more efficiently, regardless of their masses. As the gas supply is depleted, accretion onto the central black hole slows down, and AGN feedback transitions from  a quasar to a radiative feedback mode around $z=2$. Due to the different radiative efficiencies employed for the two feedback modes, \( \rm \epsilon_{f} = 0.15 \) for the quasar mode at \( \chi \geq 0.01  \) and \( \rm \epsilon_{f} = 1.0  \) for the radio mode at \( \chi < 0.01  \), (see Section \ref{subsec:AGN_feedback}), a larger percentage of accretion energy is converted into feedback at redshifts below \(z<2 \), but nevertheless, the amount of energy available for feedback declines. 

Summarising the impact of all three effects, we conclude that the decreasing AGN power for a galaxy of a given stellar mass is driven by the cumulative effect of a proportionally larger central SMBH and the decreasing gas supply in the galaxy, for which the increasing efficiency of the feedback mode at $z<2$ is unable to compensate. Not only do existing BHs of a given mass accrete less efficiently in the gas poor galaxies at $z<2$, such that ${\rm \dot{M}_{\rm BH}(M_{SMBH}}, z>2) > { \rm \dot{M}_{ BH} ( M_{ SMBH}},z<2) $ for all $\rm M_{SMBH}$, but any galaxy of a given stellar mass also hosts a significantly smaller black hole than its equivalent counterpart at $z>2$, ie ${\rm M_{SMBH}/M_* }(z<2)< {  \rm M_{SMBH}/M_* } (z>2)$. The two effects combine together to produce the redshift evolution of AGN power seen in Figure \ref{fig:AGNpower_mass}, despite the shift to a more efficient feedback mode around $z=2$.   

\begin{figure}
\centering
\includegraphics[width=0.5\textwidth]{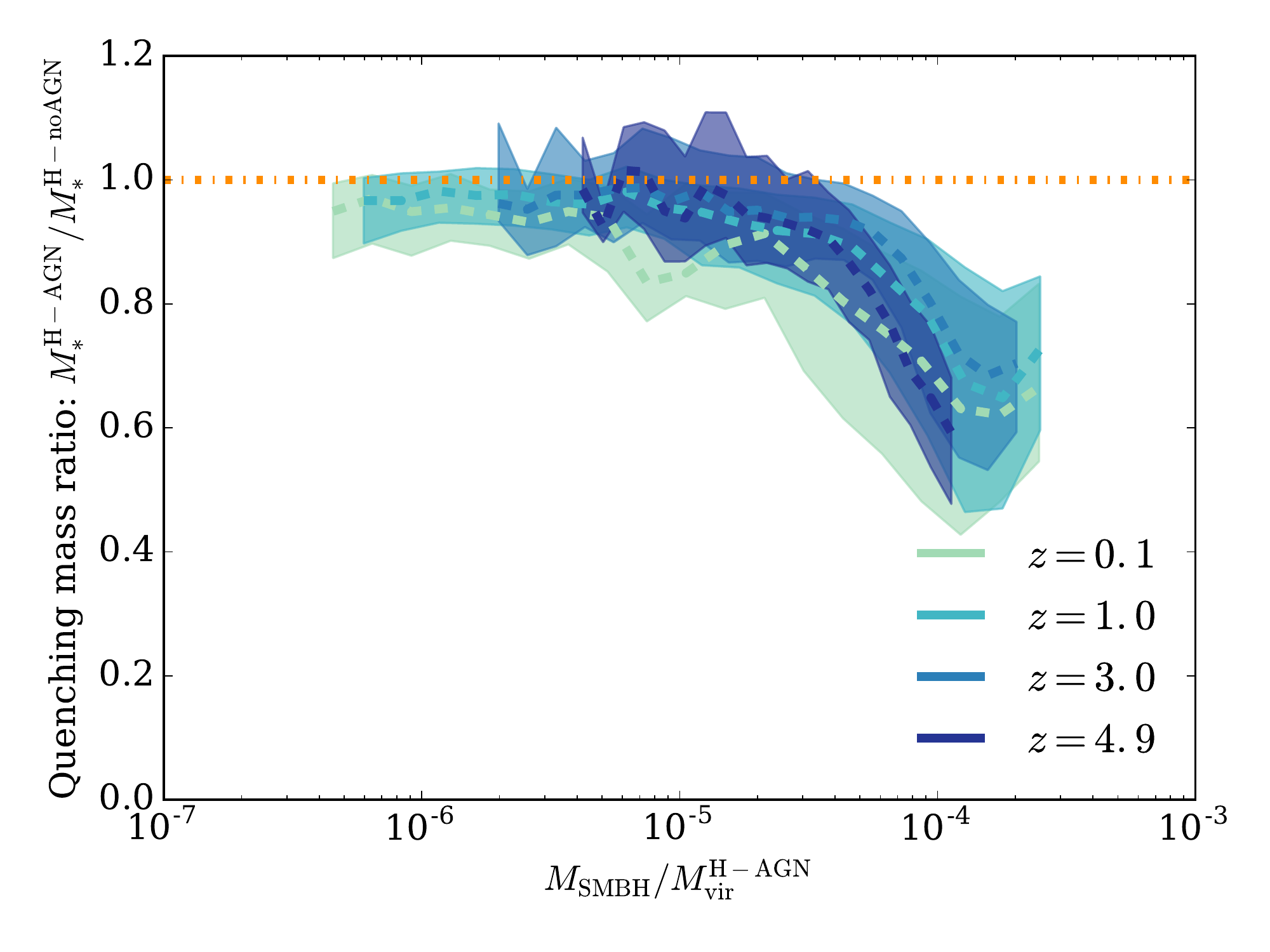}
\caption{Stellar mass ratio between H-AGN and H-noAGN galaxy twins against ratio of central SMBH mass to virial mass of the host (sub)halo, instead of stellar mass as in Fig. \ref{fig:ratios_mass_quartile}. The redshift evolution is erased in that case because the feedback energy deposited is directly connected to the central BH mass (see text for detail).}
\label{fig:ratios_BHmass}
\end{figure}

How efficiently an AGN of a given power is able to remove gas from a galaxy is dependent on the depth of the gravitational potential it has to overcome in the process. Repeating the analysis of the quenching mass ratios, but plotting it against $\rm M_{\rm SMBH} / M^{H-AGN}_{vir}$ instead of the galaxy stellar mass, shows that the redshift evolution in the quenching mass ratio is truly driven by the evolution of the SMBH population and its feedback power (Fig. \ref{fig:ratios_BHmass}, in comparison to Fig. \ref{fig:ratios_mass_quartile}). Indeed, whilst the relation scatter augments, the different redshift curves now overlap within the quartile error bars: the redshift dependence in Fig. \ref{fig:ratios_mass_quartile} has been erased. In other words, independently of redshift, BHs with a black hole to halo mass ratio of less than $M_{SMBH} / M^{H-AGN}_{vir} \leq 4 \times 10^{-5} $ quench their host galaxy by less than 20 \%, whereas BHs with ratios only a factor 2-3 larger than that suppress their host galaxy stellar masses by up to $50 \%$. 

Although there exists a clear transition of the sample from one AGN feedback mode to the other, no significant difference was found when analysing the quenching mass ratio (such as in Fig. \ref{fig:ratios_BHmass}) by splitting the sample into quasar or radio mode galaxies. All results shown here can be reproduced by assuming that the entire sample can be found in quasar mode at redshifts \(z>2 \), and in radio mode otherwise. The only notable discrepancy between the simulation data and this simplified model is that the scatter is somewhat reduced, which is expected as objects found in the opposite feedback mode to the majority of the population are statistical outliers. However, this analysis is based on an instantaneous measure of feedback mode at a specific redshift, and does not capture the accretion history of a particular object. We defer a more careful analysis of the evolution of the AGN sample, together with an analysis of the timescales on which quenching occurs in individual galaxies, to future work.

\section{AGN feedback \& gas flows} 
\label{sec:quenching_reasons}

Fig. \ref{fig:BHmass_mass} shows that SMBHs make up much less than 1\% of the mass of their host galaxy, so the mass of baryons accreted by the SMBHs is negligible compared to the
reduction in stellar mass caused by quenching. The effect of AGN feedback on the cold gas supply of the galaxy must therefore be profound, to suppress star formation by up to an order of magnitude over the evolution of the galaxy. There are three possible channels through which AGN feedback can affect the gas content of the galaxy: (i) it can drive powerful outflows, emptying the reservoir of gas available in the ISM of the galaxy; (ii) it can prevent cosmic inflows from replenishing the gas supply in the galaxy or (iii) it can heat existing gas of the ISM and circum-galactic medium (CGM) to prevent cooling flows and the associated star formation. In this section, we investigate the relative importance of these three feedback channels.

\subsection{The evolution of the baryon content}
\begin{figure}
\centering
\includegraphics[width=0.5\textwidth]{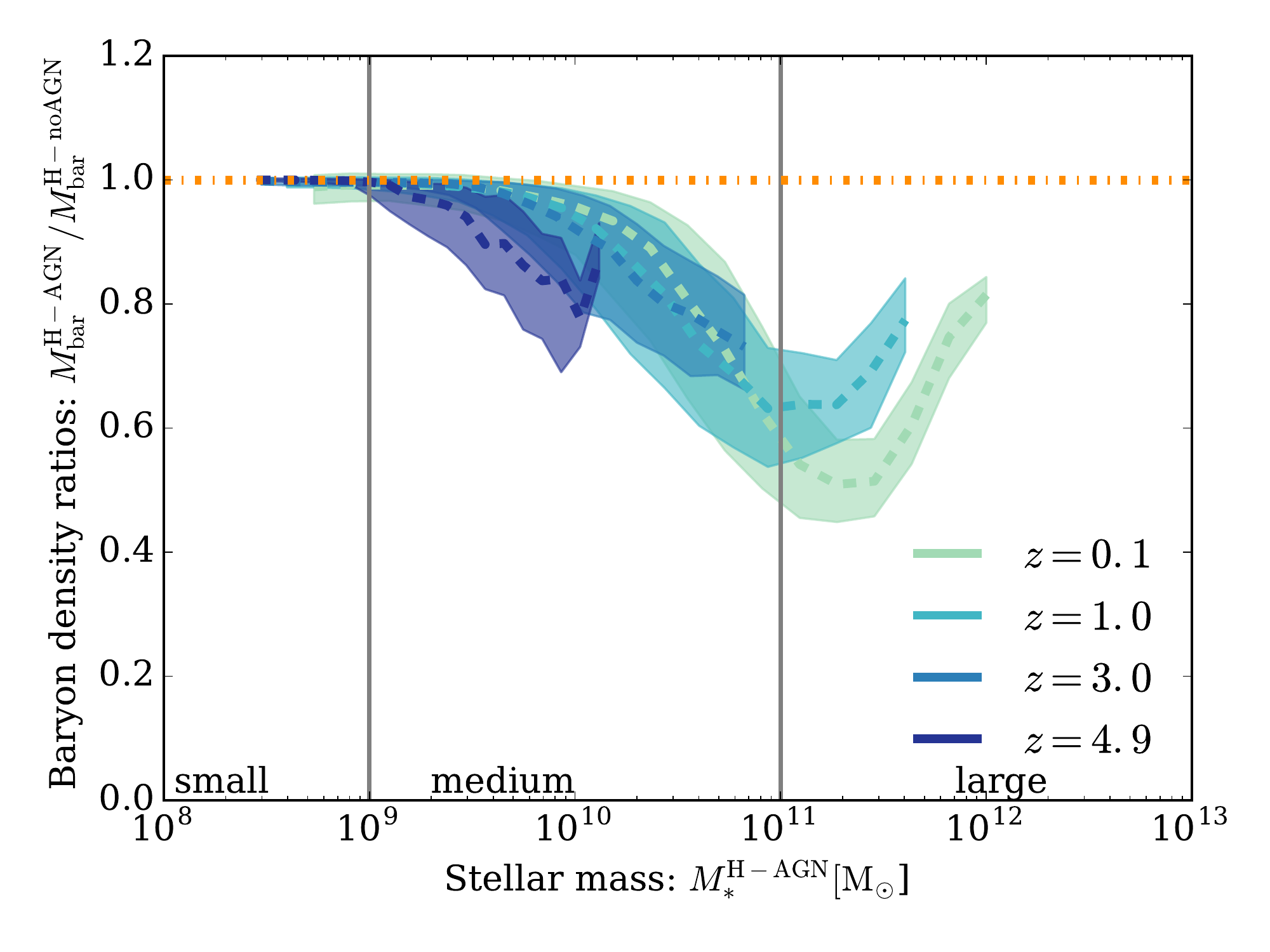}
\caption{ Ratio between the average baryon density
\( \rm \overline{\rho}_{bar}^{H-AGN} / \overline{\rho}_{bar}^{H-noAGN}   \) within the virial radius of the galaxy
host halo twins: \( \rm \overline{\rho}_{bar} = (M_{*} +M_{SMBH} +M_{gas})/(\frac{4\pi}{3}R_{vir}^3) \). 
The dotted line represents identical baryon mass in both host halos, and the shaded regions show the quartile ranges of the sample. 
AGN feedback partially acts on star formation by reducing the
total baryon mass in the halo, particularly for galaxies with stellar masses
around \(\rm M_*^{H-AGN} = 10^{11} M_\odot  \) in H-AGN.}
\label{fig:baryon-ratios_mass}
\end{figure}

Should AGN feedback primarily suppress star formation through heating the existing gas in the halo, one would expect twinned halos to have the same total baryon mass, with that in H-AGN showing a much higher gas fraction as less gas is being turned into stars. Fig. \ref{fig:baryon-ratios_mass} shows that AGN feedback directly lowers the baryon (gas + stars + BH) content embedded within the virial radius of DM halos: the average baryon density ratio versus galaxy stellar mass relation follows a shape reminiscent of that of the quenching mass ratio previously discussed. Unfortunately all three major channels, through which AGN feedback is expected to affect star formation, lower the baryon density of the galaxy. Boosted outflows drive existing gas out of the galaxy, slowed down inflows prevent accretion in the first place, and heating causes the gas to expand, lowering the average density. We compare the average baryon density, as opposed to the total mass within the halo, to correct for the small differences in halo mass shown in Fig. \ref{fig:ratios_DM}, which translate into a difference in virial radius, and therefore a difference in the volume over which the gas mass in the halo is measured. 

Two features stand out in comparison to the quenching mass ratio. First, the average baryon density within the halo of small galaxies is unaffected by AGN feedback at all redshifts. These galaxies do however show a reduced stellar mass, particularly at redshift \(z=5 \), where their stellar mass shows a median reduction of 10\%, as seen in Fig. \ref{fig:ratios_mass_quartile}. This suggests that feedback affects the star formation efficiency of these galaxies more than it alters their gas supply. Efficiency is reduced either by locally heating the gas, redistributing the gas within the halo, or by destroying dense, star-forming clumps, but not by driving outflows or preventing gas inflows through the halo virial sphere.  Secondly, for large galaxies with stellar masses above \(\rm M_* > 10^{11} M_\odot  \), the baryon content in both simulations becomes increasingly comparable again with increasing mass, despite the fact that these galaxies see a reduction in their stellar mass of around 80\% for redshifts between \(z=1 \) and $z=0.1$. In this case, the deepening gravitational potential of the halo makes it difficult for the AGN to affect gas flows at the halo virial radius. However a much smaller fraction of the existing gas is converted into stars, because of effects (gas heating / redistribution) similar to those which plague small galaxies. These translate into the significantly reduced galaxy masses presented earlier in Fig. \ref{fig:ratios_mass_quartile}.  For medium galaxies at all redshifts, AGN feedback acts through depleting the gas reservoir at the halo scale, which reduces the supply of gas available for star formation. This means AGN feedback also directly influence the inflows and/or the outflows of the galaxy. We leave a detailed analysis of the interstellar and intergalactic medium under AGN feedback to future work (Beckmann 2017, in prep.).

\subsection{The effect on inflows and outflows}
\label{subsec:flows}

\begin{figure*}
	\centering
	\begin{tabular}{c}
		\includegraphics[width=\textwidth]{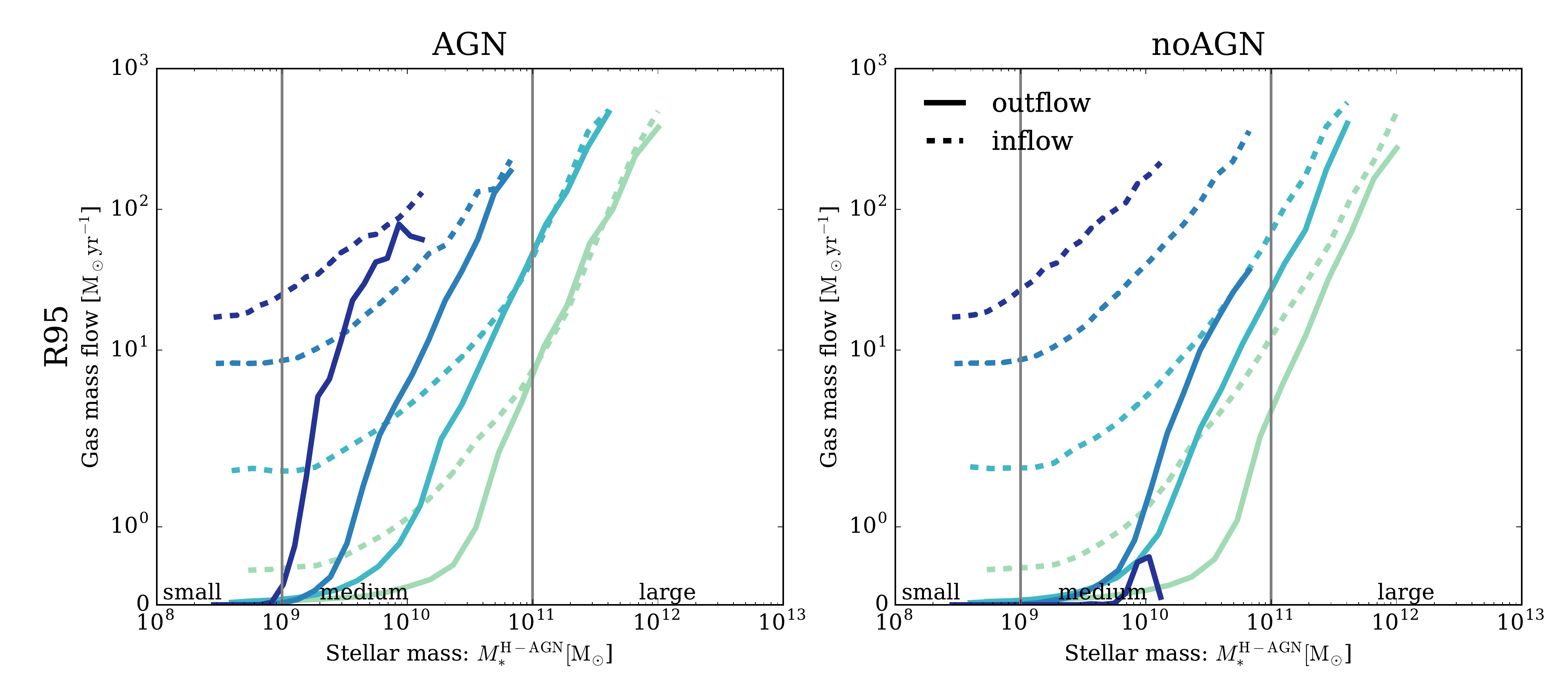} \\
		\includegraphics[width=\textwidth]{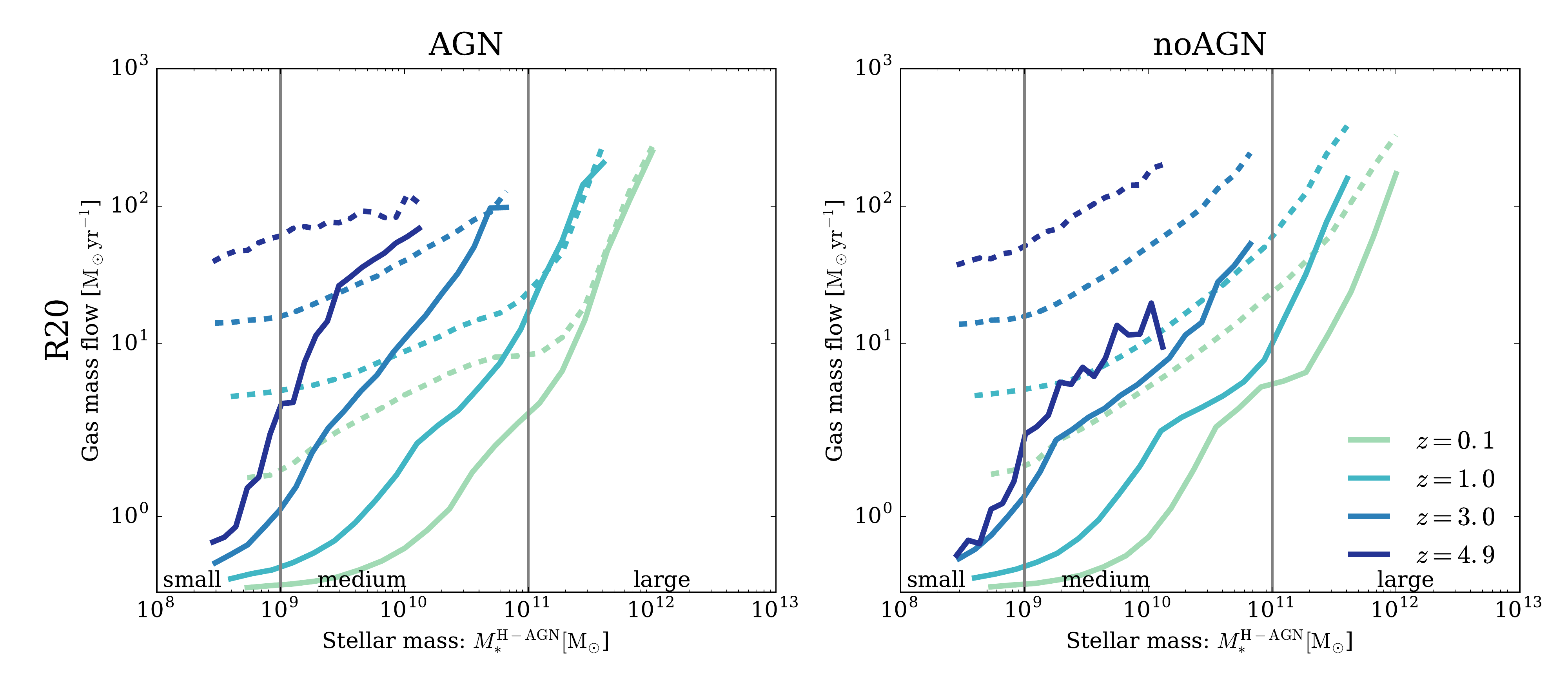} 
	\end{tabular}
	\caption{Gas inflows (dashed lines) and outflows (solid lines) at halo (R95, measured at \(\rm R=0.95 R_{vir}  \)) and galaxy (R20, measured at \(\rm R=0.2 R_{\rm vir}  \)) scales for both H-AGN and H-noAGN, for a range of redshifts. To simplify the comparison with other plots, flow values for both H-AGN and H-noAGN are plotted against H-AGN galaxy stellar masses. Note that both outflows and inflows represent the median value for a given mass bin, and do not necessarily belong to the same object in H-AGN and H-noAGN. The overall inflows are reduced in the presence of AGN feedback, which also drives outflows at halo scales for medium size galaxies.}
	\label{fig:flows}
\end{figure*}
	
There are two ways to decrease the total baryon mass of a galaxy: by reducing inflows or by boosting outflows. In this work, we measure flows at two different radii: halo scales, also called R95, which correspond to a radius of \(\rm R95=0.95 \times R_{vir}  \), and galaxy scales, also called R20, which correspond to a radius of  \(\rm R20=0.2 \times R_{vir}  \). Flows are measured through spherical surfaces located at these radii, centred on the halo. Flow masses are calculated for all cells within a narrow shell, centred on the radius in question, where \(\rm \dot{M}_{gas}=\sum_i \rho \Delta x_i^3 \bar v_i \cdot \bar r_i/\omega\), where  \(\rho\) is the gas density, \(\rm \Delta x\) is the cell size, \(\rm \bar v_i\) is the gas velocity, \( \rm \bar r_i\) is the unit vector of the cell centre relative to the halo centre and \( \rm \omega = 2kpc \) is the width of the shell. \(\rm M_{outflow}\) includes all cells with \(\rm \bar v_i \cdot \bar r_i > 0\) and \(\rm M_{inflow}\) includes all cells for which \(\rm \bar v_i \cdot \bar r_i \leq 0\).

A first comparative look at the flow patterns for galaxies in H-AGN and H-noAGN (see Fig. \ref{fig:flows}) suggest that AGN feedback drives outflows in medium and large galaxies, particularly at halo scales. There are also some large scale pseudo flows present for large galaxies. These pseudo flows appear in Fig. \ref{fig:flows} because the algorithm used to extract the absolute flow values presented here assumes the halo can be accurately represented by a sphere. However, if the halo is non-spherical, pseudo flows are created. When rotating a non-spherical object through a spherical surface across which absolute mass flows are measured, parts of the object passing out of the sphere will register as outflows, while parts passing in will register as inflows. However, these contributions are not mass flows in the common sense, and cancel out when calculating net mass flows. 
 
\begin{figure}
	\centering
		\includegraphics[width=0.5\textwidth]{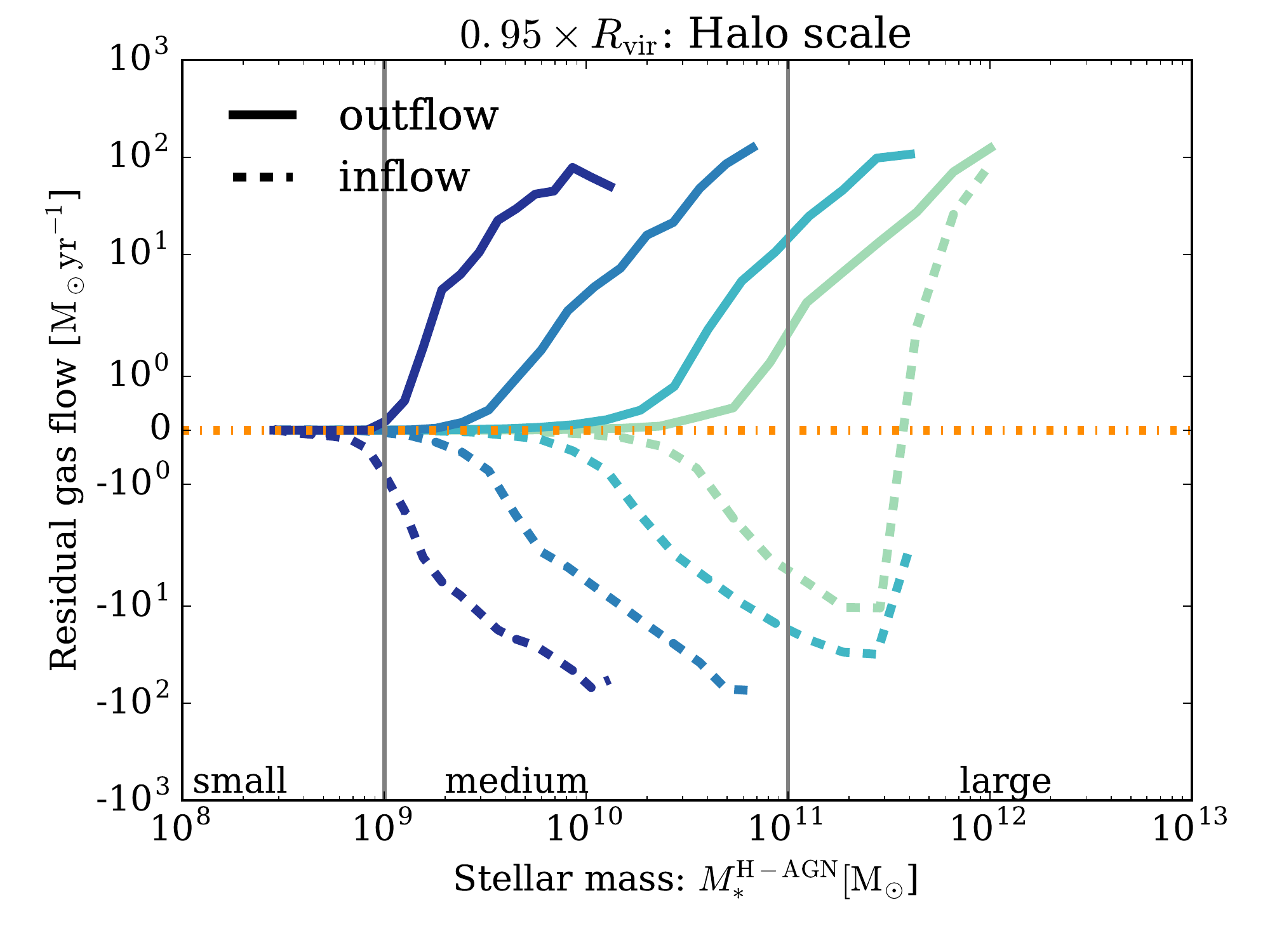}
		\includegraphics[width=0.5\textwidth]{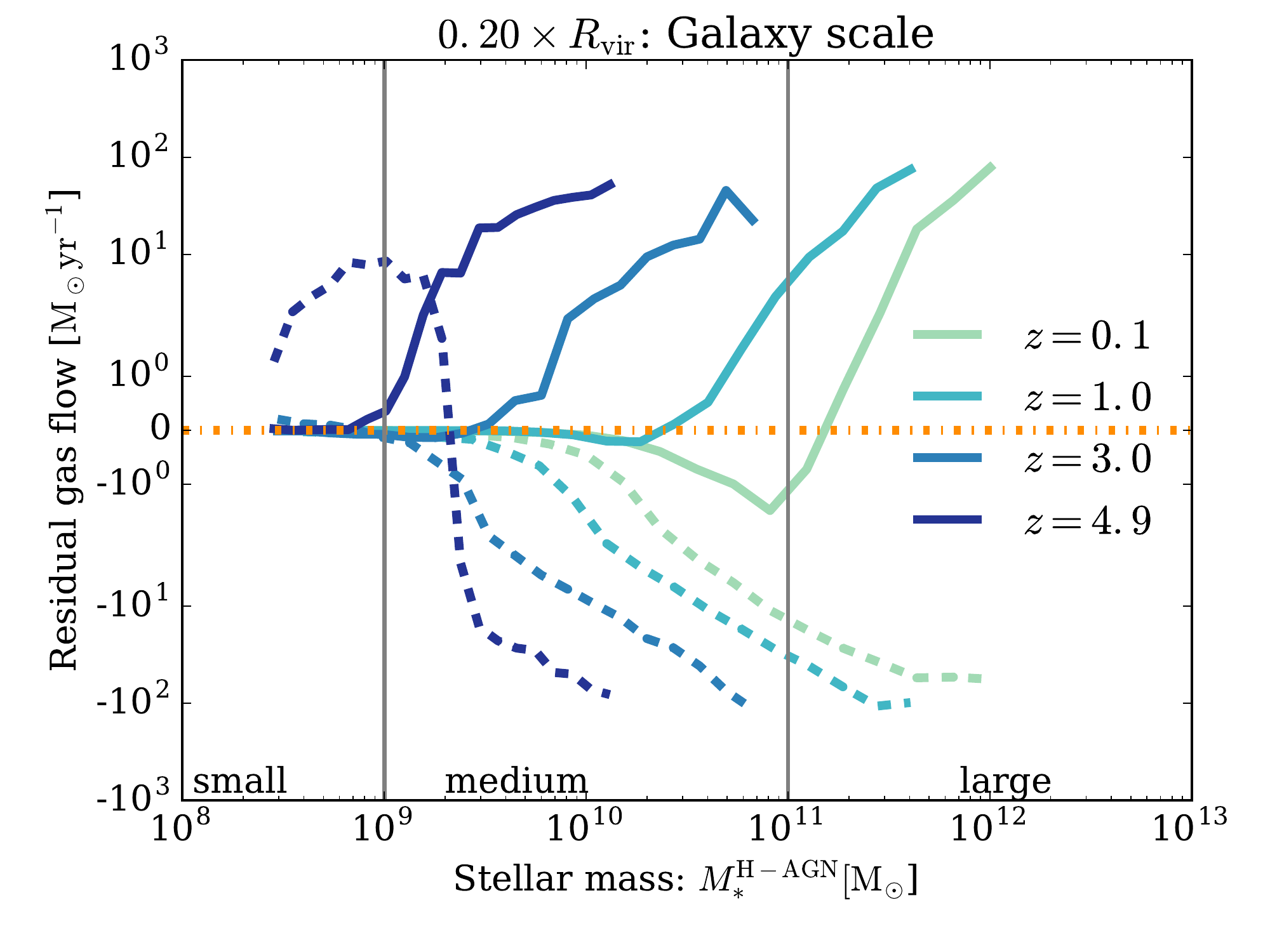}
	\caption{Residual gas flows  \(\rm \dot{M}_{gas}^{residual}=\dot{M}_{gas}^{H-AGN}-\dot{M}_{gas}^{H-noAGN}\), plotted against the twin's galaxy stellar mass in H-AGN. Data is presented for flows at two different distances from galaxies, as indicated above each panel, and a range of redshifts. Higher mass flow rates in H-AGN appear above the dash-dotted line, which denotes identical flows in both simulations, and lower mass flow rates sit below this line. Note that both outflows and inflows represent the median value for that mass bin, and do not necessarily belong to the same twin. In general, AGN feedback causes a boost in outflows, and an approximately equal but opposite reduction in inflows.}
	\label{fig:excess_flows}
\end{figure}

Small galaxies undergo no outflows at halo scales, with or without AGN feedback, which matches the conclusion from Fig. \ref{fig:baryon-ratios_mass} that the baryon mass of their halos is identical in H-AGN and H-noAGN. AGN feedback reduces inflows for medium and large galaxies, at both halo and galaxy scales, but the effect is more pronounced at the latter. A more quantitative analysis of the outflows driven by AGN feedback is presented in Fig. \ref{fig:excess_flows}, where residual flow values for the two simulations are plotted. Residual flows are defined as the mass flow rates in H-AGN relative to those of their twin galaxies in H-noAGN, i.e  \(\rm \dot{M}^{\rm residual}_{\rm gas} = \dot{M}^{\rm H-AGN}_{\rm gas}
- \dot{M}^{\rm H-noAGN}_{\rm gas}  \). This approach has the advantage of isolating the effect of AGN feedback and subtracting out any effects present in both simulations, such as the supernova driven outflows for small and medium galaxies at halo scales, and the pseudo flows seen for large objects at both radii. 

\begin{figure}
	\centering
		\includegraphics[width=0.5\textwidth]{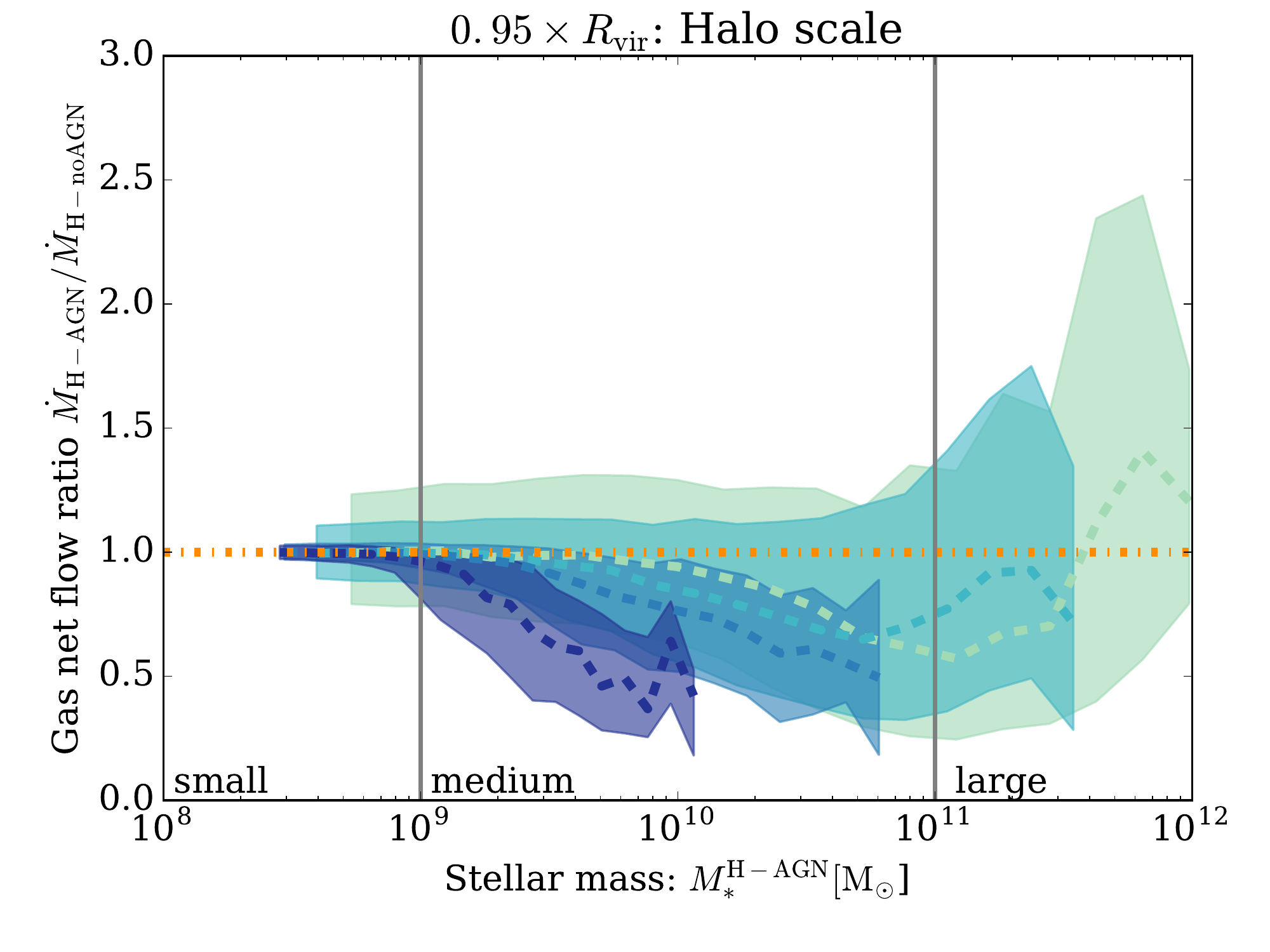}
		\includegraphics[width=0.5\textwidth]{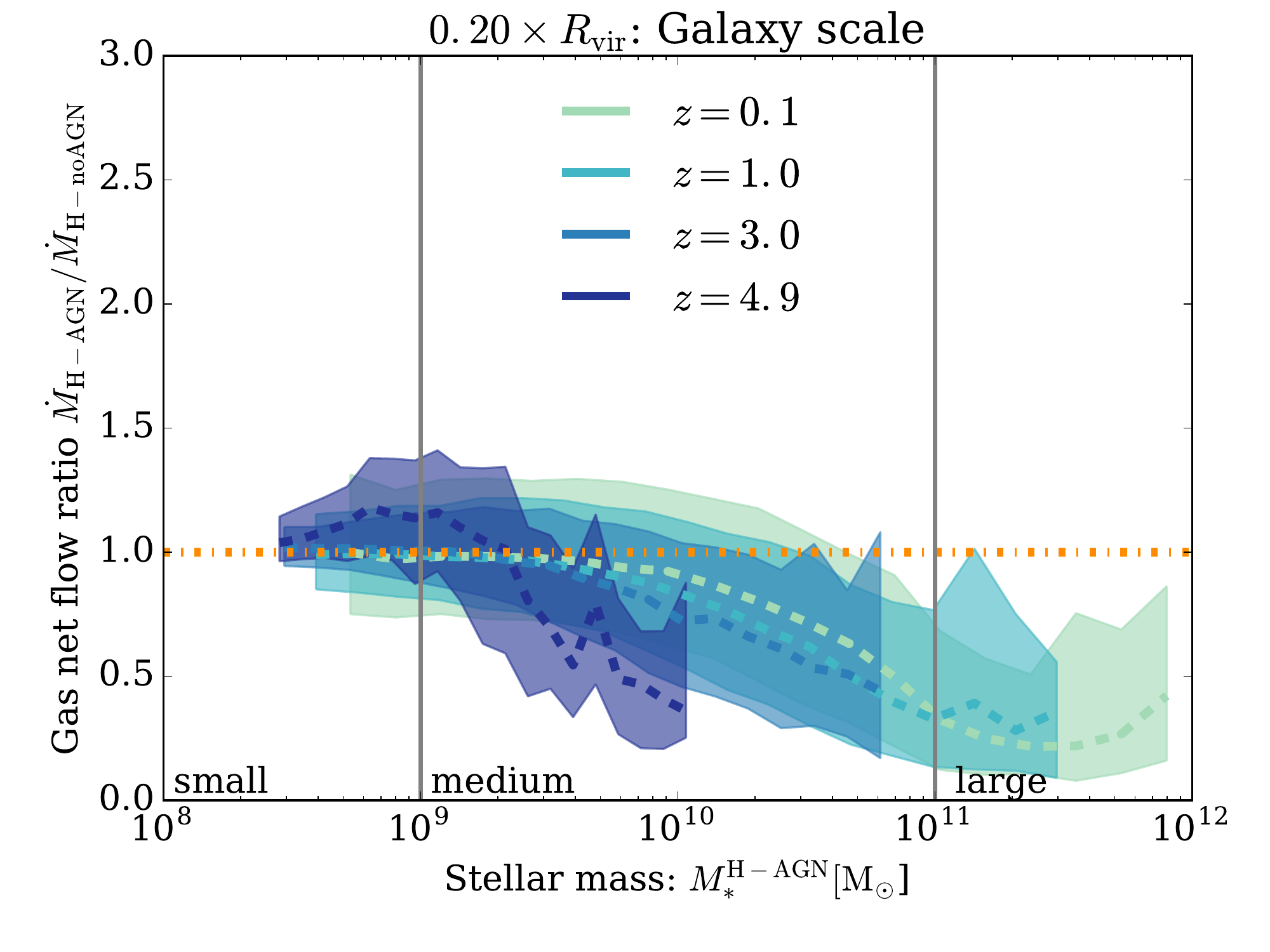}
	\caption{Ratio of net flow rates at halo and galaxy scales as a function of galaxy stellar masses in H-AGN, \(\rm M_*^{H-AGN}\). The dash-dotted horizontal line at a ratio of \(1.0 \) corresponds to identical net flows in both H-AGN and H-noAGN. Data is presented for flows at two different distances from galaxies, as indicated above each panel, and at a range of redshifts. AGN feedback has no effect on flows in small galaxies, produces a reduction in median net flow rates of up to 60\% for medium size galaxies, and drives a bursty flow pattern leading to increased net flows at halo scales for large galaxies. Positive net flows are defined to be falling into the object.}
	\label{fig:flowratios}
\end{figure}

The residual gas flows shown in Fig. \ref{fig:excess_flows} demonstrate that AGN feedback has an approximately equal and opposite effect on outflows and inflows. Galaxies with stellar masses \(\rm M_*^{H-AGN} \leq 10^{11} M_\odot  \) see a similar amount of gas carried away by AGN driven outflows as that depleted in inflows. Apart for small galaxies at high redshift ($z=5$), where AGN feedback seems able to heat up the gas in the vicinity of the galaxy, causing a gas pile up which triggers larger inflows in H-AGN, the differences between flows at halo and galaxy scales are rather modest. There is a weak trend for boosted outflows to be less dominant at galaxy scales (especially at $z \leq 1$) than at halo scales, and conversely for the suppression of inflows to be more relevant on small scales (especially for galaxies which are less massive than \(\rm M_*^{H-AGN} \leq 10^{10} M_\odot  \)), but overall material is neither being significantly swept up in the halo and kicked out, nor preferentially deposited there. Rather, these two effects combine to reduce the net median inflow for medium galaxies into both the halo and the galaxy by up to 60\% for the most massive objects, as the direct comparison of net inflows for twins in Fig. \ref{fig:flowratios} shows. This results in the reduction in baryon mass seen for the same galaxies in Fig. \ref{fig:baryon-ratios_mass}. In agreement with the same figure, small galaxies show no change in net flows at halo scales, which matches their identical baryon mass in the presence and absence of AGN feedback.  These results are in agreement with work by \citet{Dubois2013}, who compare high resolution zoom simulations with and without AGN feedback of a single galaxy of $M_*(z=6)=6.2\times10^9 M_\odot$. The authors find strong evidence for the fact that the AGN significantly heat the gas at halo scales, driving a hot super-wind and destroying cold flows. \citet{Pontzen2016} also report AGN boosted outflows in their hydrodynamical merger simulation, and emphasize that long term quenching requires the inflows to be suppressed.

For large galaxies, the situation is different, particularly at low redshift. At halo scales, the simulation with AGN feedback actually shows boosted inflows carrying an amount of mass similar to that in boosted outflows. This means the outflows are being recycled, as AGN feedback becomes unable to gravitationally unbind the gas from the halo. It is important to note that the values plotted here represent the median value for a given mass bin, so the outflow and inflow values do not necessarily belong to the same object. It is therefore not necessarily correct that the two curves cancel out to produce no change in the net flow. Indeed, a comparative analysis of net inflows for each twin across H-AGN and H-noAGN (Fig. \ref{fig:flowratios}) shows that at halo scales, the overall inflow is boosted by up to 50\% for the most massive galaxies in the presence of AGN feedback.

\begin{figure*}
	\centering
	\includegraphics[width=\textwidth]{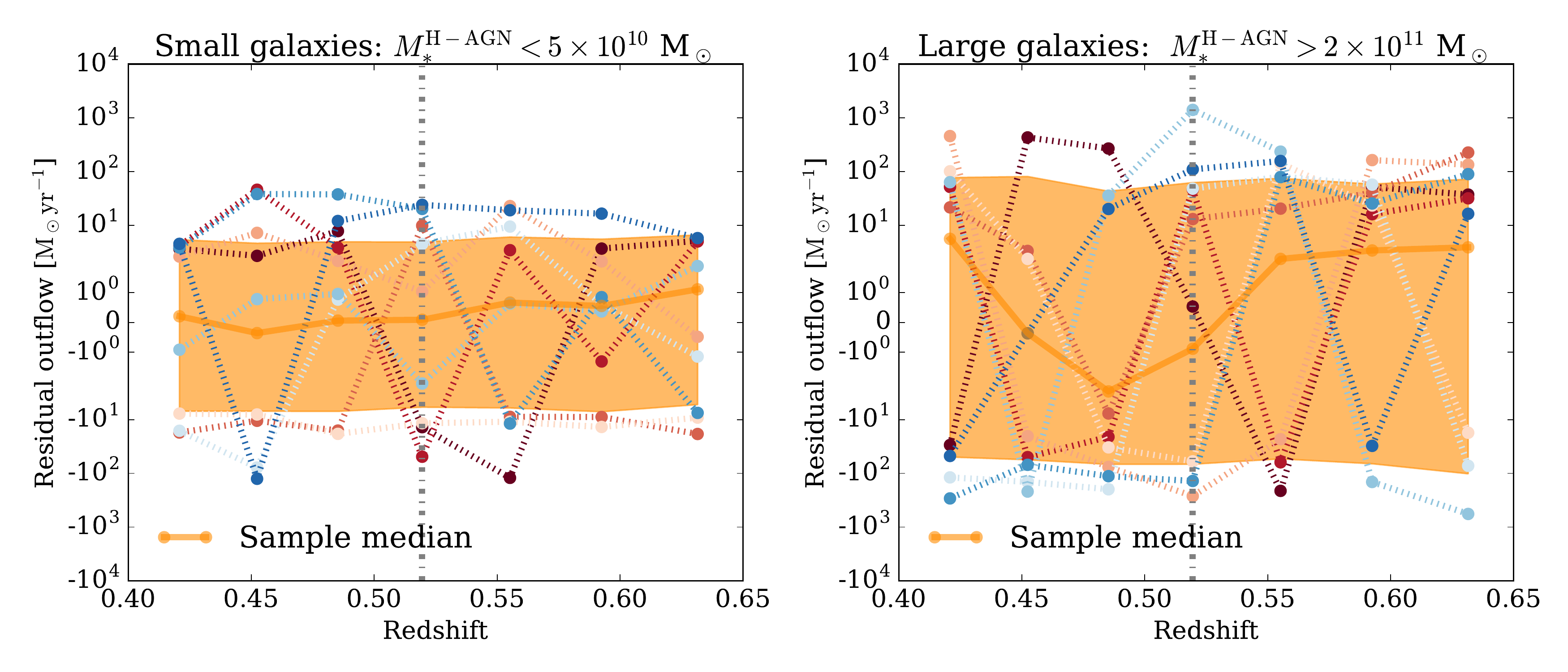} 	
	\caption{ Residual outflows due to AGN feedback at galaxy scales, \(\rm R20=0.2 \times R_{vir} \), plotted for 10 randomly selected galaxies within the small and large galaxy mass bins, between redshifts  $0.64 > z > 0.42$. The solid line represents the median sample output, with the shaded
region covering quartile error ranges. The vertical dotted line represents redshift $z = 0.52$, for comparison with Fig. \ref{fig:excess_flows}. This shows that outflows from large galaxies (right panel) with stellar masses \(\rm M_*^{H-AGN} > 2 \times 10^{11} M_\odot \) at redshifts around $z = 0.52$ (dotted vertical line) vary on very short timescales, so the median outflow value, (solid line, with quartile error ranges represented by the shaded regions), is very sensitive to the exact time at which it is sampled. By comparison, the outflows for small galaxies with stellar masses \(\rm M_*^{H-AGN} < 5 \times 10^{10} M_\odot \)  (left panel) vary much less on these timescales, and the median value is less sensitive to variations in time.}
\label{fig:dutycycle}
\end{figure*}

At galaxy scales, the gas flow patterns for the most massive objects, \( \rm M_*^{H-AGN} >  \times 10^{11} M_\odot \) at $z=0.1$), become harder to predict. In these galaxies, AGNs fall into maintenance feedback mode (see Section \ref{subsec:BHevolution}) at redshifts below the peak of star formation, \(z=2 \). This produces very bursty outflows, as SMBHs go through cycles of being fed, which triggers strong feedback episodes. The latter drive out the gas, starving the black hole, and the feedback abates until enough gas becomes available again for the SMBH to go through another accretion event. These cycles take place on timescales much shorter than the time interval between the redshift outputs we are considering in this work. Furthermore, the number of such galaxies is quite limited ($\approx$ 1000). We illustrate the impact this has on our results in Fig. \ref{fig:dutycycle}, which shows that at redshift \(z=0.5 \), the median outflow depends quite sensitively on the exact point in time at which the distribution of galaxies is sampled. As the SMBHs cycle rapidly through a wide variety of active and quiescent states, the distribution of residual outflows spans several orders of magnitude. In comparison, the outflows produced by the larger population of smaller galaxies are more steady on similar timescales, and therefore their sampling is more robust (see left panel of Fig. \ref{fig:dutycycle}). We would like to point out that the variation timescales of the large scale outflows studied here do not necessarily reflect the duty cycle of the SMBH, as each burst can be driven by a series of feedback events. We postpone a more detailed analysis of the SMBH duty cycles in the simulation to future work (Beckmann 2017 et al, in prep). 

It is interesting to note that the lack of excess outflows for low mass galaxies (\( \rm  M^{H-AGN}_* < 5 \times 10^{10} \  M_\odot \) at $z=0.1$) in Fig. \ref{fig:excess_flows} does not mean that the outflows for any given matching galaxy pair in H-AGN and H-noAGN are identical. As the panel on small galaxies in Fig. \ref{fig:dutycycle} demonstrates, individual objects show a variety of residual outflows (and inflows). The build-up of small differences in galaxy properties not necessarily induced by AGN feedback (e.g. stochastic star formation algorithm, seeding and growth of the central SMBH) can result in temporarily diverging residual outflow histories at low redshift. In other words, even though the precise amount of residual outflow from any specific twinned pair of galaxies does depend on the timescale of the outflow and so is sensitive to the redshift at which it is measured, the lack of any marked systematic difference in the gas flow pattern due to AGN feedback registers as a median residual outflow of zero for the whole sample. 

Another note of caution concerns the residual flows in Fig. \ref{fig:excess_flows}, which likely underestimate the effect of AGN on inflows and outflows.  Particularly for massive galaxies at low redshift, where the H-noAGN twin has a stellar mass \(\sim 5 \times \) that of the H-AGN one, the H-noAGN twin has stronger stellar flows that obscure some of the AGN driven effects when calculating residual gas flows as \(\rm \dot{M}_{gas}^{residual}=\dot{M}_{gas}^{H-AGN}-\dot{M}_{gas}^{H-noAGN}\).  Finally, we come back to the asymmetry between residual inflows and outflows with respect to the zero residual line, which is stronger at galaxy than halo scales. In light of the previous remarks about outflow timescales, we can safely attribute this difference as meaning that AGN feedback does preferentially suppress inflows in the vicinity of all galaxies rather than eject gas from them, except perhaps at the very high mass end of the galaxy stellar mass function, \(\rm M_{*}^{H-AGN} > \times 10^{11} M_\odot\) at low redshifts ($z < 0.1$).  

Overall, flow patterns due to AGN feedback, made up in roughly equal parts of boosted outflows and reduced inflows at halo scales, comfortably explain the non-linear distribution of baryon masses plotted in Fig. \ref{fig:baryon-ratios_mass}, which combines with a reduced star formation efficiency across all galaxy mass bins to produce the quenching mass ratios shown in Fig. \ref{fig:ratios_mass_quartile}.

\section{Discussion \& Conclusions}
\label{sec:conclusions}

We have isolated the effect of AGN feedback on stellar quenching in massive galaxies by comparing two cosmological simulations, H-AGN and H-noAGN, which were run with and without AGN feedback respectively. More specifically, by twinning individual DM halos and galaxies across the two simulations, we have been able to quantify the effect of feedback on individual objects throughout cosmic time. In agreement with a large body of previous work \citep[such as][among others]{Springel2005,Bower2006,Sijacki2007,Ciotti2007,Cattaneo2007,Dubois2013,Gene2014,Schaye2014,Pontzen2016} our results show that AGN feedback is instrumental in quenching the massive end of the GSMF. Whilst the stellar mass of galaxies without AGN feedback closely follows predictions based on the assumption that all baryons contained in dark matter halos end up forming stars, galaxies subject to the influence of AGN feedback end up with masses distributed according to a GSMF that shows a characteristic exponential steepening at the high mass end, in line with observations \citep{Bielby2012,Fontana2004,Moustakas2013,Tomczak2013,Gonzalez2011,Song2015}.

The importance of AGN feedback has been emphasised in all recent large-scale simulations of galaxy evolution, but the results differ in the details. Similar to results presented here, galaxies in the MassiveBlackII simulation exhibit signs of relatively strong quenching early on but then see a reduction in the impact of AGN feedback at lower redshifts \citep{Khandai2015}. A similar issue is reported by Illustris, who found that despite aggressive AGN feedback that produces unrealistically low gas fractions in DM halos at low redshift, star formation is not suppressed strongly enough and the simulation overproduces massive galaxies \citep{Gene2014,Vogelsberger2014,Vogelsberger2014a}. As opposed to the dual AGN feedback model used in H-AGN and Illustris, the EAGLE simulation only employs a single feedback mode, tuned to reproduce the GSMF at $z=0$. They do also see good agreement with the GSMF at redshift $z>2$, which supports our conclusion that the shift in feedback mode plays a subordinate role in the importance of AGN feedback \citep{Schaye2014}. Overall, regardless of the hydrodynamics scheme employed and the detail of the subgrid model implementation of AGN feedback, all four recent large-scale cosmological simulations of galaxy evolution agree that AGN feedback must play a crucial role in regulating the evolution of massive galaxies.  

A closer comparison of the stellar mass for individual objects in H-AGN and H-noAGN reveals a non-linear dependence of quenching mass ratio on mass, with the most massive galaxies being the most strongly quenched, and the smallest galaxies mostly unaffected by AGN feedback. This leads to a characteristic shape for the mass ratio \(\rm M_*^{H-AGN}/M_*^{H-noAGN} \), which shows a linear dependence on \(\rm \log(M_*^{H-AGN})  \) for medium sized galaxies with  \( \rm 10^9 M_\odot \lesssim M_*^{H-AGN} \lesssim 10^{11} M_\odot  \) (the exact values depend on redshift), but tails off at both low and high mass ends. The most massive galaxies, with \(\rm M_* > 10^{11} M_\odot  \) are most strongly quenched and contain only 20\% of the stellar mass in the presence of AGN feedback at $z=0.1$, in comparison to the case without feedback. 

We also find a significant redshift evolution for the smallest galaxy mass to be affected by AGN feedback, with smaller galaxies being more quenched at higher redshift. This transition mass at which AGN feedback becomes important, evolves from $2.24 \times 10^{10} M_\odot$ at $z=0.1$ to $1.48 \times 10^{9} M_\odot$ at $z=4.9$. Such a redshift dependence seems in good agreement with observations \citep{Baldry2004,Peng2010} but systematically leads to values at high redshift which are lower than these reported by a range of hydrodynamics simulations \citep{DiMatteo2011,Keller2016,Pontzen2016,Bower2017}. We argue that these discrepancies reflect the fact that quenching is very likely a cumulative process that builds up over the entire history of the galaxy, not just a one-off event that shuts down star formation forever when the galaxy reaches a particular stellar mass. If correct, the consequence is that differences induced by AGN feedback in star formation rates can be small, particularly for objects near the transition mass, the effect of such feedback only becoming apparent for galaxies once its integral represents a significant fraction of their total stellar mass. 

In our case, this evolution is caused by the median central black hole mass of galaxies of a given stellar mass, $M_{\rm SMBH}/M_*$, increasing with redshift by a factor of a few between $z=0$ and $z=5$, a result for which there exists observational support \citep{Merloni2010,Decarli2010}. However, it has been suggested that the observation trend might be biased \citep{Lauer2007,Volonteri2011,Bongiorno2014}. A similar trend has been reported in other large scale simulations, such as MassiveBlackII \citep{Khandai2015} and Illustris \citep{Sijacki2015}, but not in all of them \citep{Rosas-Guevara2016}. This evolution in black hole masses combines with higher accretion rates at high redshift, due to gas-rich galaxies, such that a galaxy of the same stellar mass is subject to AGN feedback up to three orders of magnitude stronger at redshift \(z=5\) than at \(z=0.1\). We also measure a shift in feedback mode, with at least 85.5\% of AGN at redshift \(z=3 \) and above in quasar mode, compared with a maximum of 19.1\% at lower redshifts, but the increasing radiative efficiency associated with this shift is unable to offset the trend of lower $\rm M_{\rm SMBH}/M_*$ and lower accretion rates. 

A comparative analysis of the baryon content of halos reveals that AGN feedback quenches star formation through a combination of reducing the total gas supply within the halo by driving outflows, and preventing accretion of fresh gas by curbing inflows. Small galaxies with \( \rm  M_*^{H-AGN}  \leq 10^9 M_\odot \) show nearly identical baryon masses with and without AGN feedback, as the flows at halo-scales remain chiefly unaffected by feedback. Note that this is not true at galaxy scales and high redshift, where gas inflows can be somewhat enhanced by feedback for these small objects. On the other hand, medium size galaxies \( \rm 10^9 M_\odot \leq M_*^{H-AGN} \leq  10^{11} M_\odot  \) experience a significant reduction in baryon mass, caused by an approximately equal contribution from AGN-driven gas outflows and a reduction of cosmic inflows. At galaxy scales, the reduction of inflows dominates. Finally, for large galaxies (\( \rm 10^{11} M_\odot \leq M_*^{H-AGN}  \)) the baryon mass rises again, as inflows at halo scales are swelled by gas expelled by AGN feedback in the inner regions, which remains gravitationally bound and falls back into the halo. At galaxy scales, outflows for the most massive objects vary on very short timescales, as the AGN enter a bursty maintenance mode. Thus the gas mass has a tendency to increase on average (as compared to medium size galaxies), even though the stellar mass does not, given the long characteristic timescale of star formation. 

The picture that emerges seems consistent with high resolution work (20 pc instead of 1 kpc) by \citet{Dubois2013}, who use zoom simulations of individual objects with and without AGN feedback, and find that AGN drive hot super-winds which disrupt cold inflows on halo scales. The two effects combine to reduces the baryon content of galaxies by up to $30 \%$. Similar conclusions were reached by early semi-analytic work by e.g. \citet{Benson2003}, who argue that strong, hot winds are necessary to reproduce the observed luminosity function, and recent simulations by \citet{Pontzen2016} and \citet{Taylor2015} that also report AGN boosted outflows, concluding that long term quenching requires gas inflows to be suppressed. 

Looking at cosmic accretion specifically, \citet{Nelson2015} find that the simulation without feedback (neither AGN driven nor stellar) see much higher levels of smooth accretion into the galaxy than the ones with (AGN and stellar) feedback. However, contrary to the results presented here, they find no evidence for recycling of gaseous material at the halo boundary.
While all current large scale cosmological simulations include AGN feedback as in integral part of their galaxy evolution model, some authors contend that processes in massive galaxies that do not rely on AGN feedback can reproduce galaxy mass functions through e.g. cosmic quenching \citep{Feldmann2014}  and that stellar feedback super-bubble feedback can drive powerful outflows \citep{Keller2016}. \citet{Gabor2014} and \citet{Roos2014}, running idealised galaxy simulations, find AGN-driven outflows consistent with those we report here but lower impact on the star formation rate of galaxies. 

Overall, we conclude that AGN feedback provides an effective mechanism to reproduce the distribution of galaxies at and above the knee of the GSMF, over a redshift range spanning $90 \%$ of the age of the Universe. For local galaxies, AGN feedback plays an important role in stifling star formation in objects above a transition mass of $M_* \geq 2\times10^{10} M_\odot$. AGN feedback acts by reducing the stellar content of galaxies by up to 80\% (for the most massive objects) through a mixture of increased outflows and reduced inflows, combined with a decreased star formation efficiency of in situ gas.  We predict that the influence of AGN feedback should already be noticeable by redshift $z = 5$, for galaxies with relatively modest stellar masses (\( \rm M_* \approx 2 \times 10^{9} M_\odot  \)) by current epoch standards, as these objects are close to the top end of the GSMF at these redshifts. This is exciting news as the James Webb Space Telescope should be able to test this prediction in the near future.

\section*{Acknowledgements}
This work used the HPC resources of CINES (Jade supercomputer) under the allocation 2013047012 made by GENCI, and the horizon and Dirac clusters for post processing. This work is partially supported by the Spin(e) grants ANR-13-BS05-0005 of the French Agence Nationale de la Recherche and by the National Science Foundation under Grant No. NSF PHY11- 25915, and it is part of the Horizon-UK project, which used the DiRAC Complexity system, operated by the University of Leicester IT Services, which forms part of the STFC DiRAC HPC Facility (www.dirac.ac.uk ). This equipment is funded by BIS National E-Infrastructure capital grant ST/K000373/1 and  STFC DiRAC Operations grant ST/K0003259/1. DiRAC is part of the National E-Infrastructure. The research of RSB is supported by STFC, and the research of AS, MLAR and JD at Oxford is supported by the Oxford Martin School and Adrian Beecroft. NC is supported by a Beecroft postdoctoral fellowship.

%\clearpage
\bibliographystyle{mnras}
\bibliography{references} 

% Don't change these lines
\bsp	% typesetting comment
\label{lastpage}
\end{document}